# Exactly Decoupled Kalman Filtering for Multitarget State Estimation with Sensor Bias

Jianxin Yi, *Member, IEEE*, Xianrong Wan, Deshi Li

*Abstract*—The problem of multisensor multitarget state estimation in the presence of constant but unknown sensor biases is investigated. The classical approach to this problem is to augment the state vector to include the states of all the targets and the sensor biases, and then implement an augmented state Kalman filter (ASKF). In this paper, we propose a novel decoupled Kalman filtering algorithm. The decoupled Kalman filtering first processes each target in a separate branch, namely the single-target Kalman filtering branch, where the single-target states and the sensor biases are estimated. Then the bias estimate is refined by fusing the former bias estimates across all the single-target Kalman filtering branches. Finally, the refined bias estimate is fed back to each single-target Kalman filtering branch to improve the target state estimation. We prove that the proposed decoupled Kalman filtering is exactly equivalent to the ASKF in terms of the estimation results under a usual initial condition. The equivalence is also confirmed via the numerical example. Moreover, we further validate the proposed algorithm using the field experimental data of a multistatic passive radar.

## I. INTRODUCTION

MULTISENSOR systems attract a great deal of interest in many fields due to the potential to enhance performance by information fusion [1]-[4]. In practice, sensors usually have measurements biases. If not corrected, these sensor biases may drastically degrade the fusion performance. Hence bias registration becomes a prerequisite for the information fusion in multisensor systems [5]-[9].

In this paper, we discuss the problem of joint multitarget state and sensor bias estimation in multisensor systems with a centralized data processing architecture. To focus on the estimation problem, we assume that the data association has been solved in advance. Regarding the joint multitarget state and sensor bias estimation, the classical approach is to augment the state vector to include the multitarget states and the sensor biases, and then implement an augmented state Kalman filter (ASKF) [10]-[13] or other filters based on the augmented state [14], [15]. As the dimensions of the augmented measurement and augmented state increase in the ASKF, the computational complexity grows sharply with the number of targets. Moreover, various targets may start and end at different time. It leads to the frequent change of the augmented-state vector, which is not convenient for implementation. Thus, this classical structure is not flexible. A practical implementation prefers decoupled processing.

To narrow the scope of discussion, in this paper we investigate the method under the classical Kalman filtering (KF) framework as it is widely used in many systems. Regarding the corresponding decoupled processing, a two-stage approach was proposed in [16] and generalized in [17]-[19] by formulating the original ASKF into two parallel, reduced-order filters. First, it does bias-ignorant estimation for the multitarget states as if there is no bias. Then biases are obtained from the bias-ignorant estimates. The corrected target states are obtained as a linear combination of the bias-ignorant estimates and the bias estimates. This two-stage approach could be equivalent to the ASKF under certain conditions. However, its structure is still complicated in the multitarget case. When the target number changes, the filter also needs to be adjusted frequently and substantially.

The ASKF approach has also been discussed in [20]. Two approximately decoupled methods were proposed by simplifying the cross-correlation between target states and biases. The first method named macro filter has nearly optimal performance but also has a complex structure, thereby facing the flexibility problem mentioned above. The second method (called approximately decoupled KF hereafter) directly ignores the cross-correlation and then processes the target states and biases separately. It is computationally efficient but with obvious performance degradation.

Another decoupled filter for the bias estimation was proposed in [21]. The bias estimation is separated from the model via a series of algebraic operations. It is an ad hoc algorithm that only outputs the bias estimate and no target state estimate. A partitioned KF for oceanic and atmospheric data assimilation is proposed in [22]. The intention therein is to tackle the high-dimensional filtering problem. It approximates the model state by a sum of a series of independent elements. Another type of partitioned (or compressed) KF is discussed in [23] and [24]. It solves the problem where the measurement and the state transition models are decoupled between substates in a time interval but the initial estimates of the substates are not

This work was supported in part by the National Natural Science Foundation of China (61701350, 61931015), in part by the National Key R&D Program of China (2016YFB0502403), in part by the Technological Innovation Project of Hubei Province of China (2019AAA061), in part by the Postdoctoral Innovation Talent Support Program of China (BX201600117), and in part by the Fundamental Research Funds for the Central Universities (2042019kf1001).

J. Yi, X. Wan, and D. Li are with the School of Electronic Information, Wuhan University, and also with the Collaborative Innovation Center for Geospatial Technology, Wuhan, China. (corresponding author: Jianxin Yi, e-mail: jxyi@whu.edu.cn)

independent. For the multitarget state estimation with sensor biases herein, the sensor biases are coupled with all the targets in the measurement models. It does not meet the applicable condition of the partitioned KF in [22]-[24].

In this paper, our objective is to find a decoupled method that is equivalent to the ASKF and has a clear and flexible structure. To this end, we decouple the ASKF by taking each target as a basic unit. First, the states of one target and the sensor biases are jointly estimated in each single-target KF branch. The sensor bias estimate is then refined by fusing across all the single-target KF branches. Finally, the refined sensor bias estimate is fed back to each single-target KF branch to improve the state estimation. This novel decoupled KF can readily handle the case with a dynamic change of the number of targets. It also has lower computational complexity than the ASKF. Moreover, we prove that the proposed decoupled KF is exactly equivalent to the ASKF in terms of the estimation results under a usual initial condition. The proposed decoupled KF is further validated in the context of multistatic radars using both simulated and field experimental data.

The rest of the paper is organized as follows. Section II introduces the system model and the ASKF. Section III elaborates the proposed decoupled KF. Section IV proves the equivalence between the proposed decoupled KF and the ASKF. Simulations and real data evaluations are presented in Section V. Finally, conclusions are drawn in Section VI.

## II. PROBLEM STATEMENT

### A. Notations

The following notations are used in this paper.

| | |
|---|---|
| $\boldsymbol{x}_{t_n}$ | state vector of the $n$th target |
| $N$ | number of targets |
| $\boldsymbol{b}$ | sensor bias vector |
| $\boldsymbol{x}_n$ | augmented state vector including $\boldsymbol{x}_{t_n}$ and $\boldsymbol{b}$, i.e. $[\boldsymbol{x}_{t_n}^T, \boldsymbol{b}^T]^T$ |
| $\boldsymbol{x}$ | augmented state vector including $\boldsymbol{x}_{t_n}, n=1,\ldots,N$ and $\boldsymbol{b}$, i.e. $[\boldsymbol{x}_{t_1}^T,\ldots,\boldsymbol{x}_{t_N}^T,\boldsymbol{b}^T]^T$ |
| $\boldsymbol{z}_n$ | measurement vector of the $n$th target |
| $\boldsymbol{z}$ | augmented measurement, i.e. $[\boldsymbol{z}_1^T,\ldots,\boldsymbol{z}_N^T]^T$ |
| $\boldsymbol{F}_{t_n}$ | state transition matrix of the $n$th target |
| $\boldsymbol{v}_{t_n}$ | process noise of the $n$th target |
| $\boldsymbol{Q}_{t_n}$ | covariance matrix of $\boldsymbol{v}_{t_n}$ |
| $\boldsymbol{H}_{t_n}$ | state measurement matrix |
| $\boldsymbol{H}_{b_n}$ | bias measurement matrix |
| $\boldsymbol{w}_n$ | measurement noise of $\boldsymbol{z}_n$ |
| $\boldsymbol{R}_n$ | covariance matrix of $\boldsymbol{w}_n$ |
| $\hat{\boldsymbol{x}}^*$ | augmented state estimate in the ASKF |
| $\hat{\boldsymbol{x}}_{t_n}^*, \hat{\boldsymbol{b}}^*$ | subvectors of $\hat{\boldsymbol{x}}^*$ |
| $\boldsymbol{P}^*$ | covariance matrix of $\hat{\boldsymbol{x}}^*$ |
| $\boldsymbol{P}_{t_n}^*, \boldsymbol{P}_{t_{mn}}^*, \boldsymbol{P}_{tb_n}^*, \boldsymbol{P}_b^*$ | submatrices of $\boldsymbol{P}^*$ |
| $\hat{\boldsymbol{x}}_n$ | output estimate of the augmented state estimation module in the proposed decoupled KF |
| $\hat{\boldsymbol{x}}_{t_n}, \hat{\boldsymbol{b}}_n$ | subvectors of $\hat{\boldsymbol{x}}_n$ |
| $\boldsymbol{P}_n$ | covariance matrix of $\hat{\boldsymbol{x}}_n$ |
| $\boldsymbol{P}_{t_n}, \boldsymbol{P}_{tb_n}, \boldsymbol{P}_{b_n}$ | submatrices of $\boldsymbol{P}_n$ |
| $\hat{\boldsymbol{b}}_f$ | fused bias vector estimate after the bias information fusion in the proposed decoupled KF |
| $\boldsymbol{P}_{fb}$ | covariance matrix of $\hat{\boldsymbol{b}}_f$ |
| $\hat{\boldsymbol{x}}_{f_n}$ | updated estimate after the augmented state update in the proposed decoupled KF |
| $\hat{\boldsymbol{x}}_{ft_n}$ | subvector of $\hat{\boldsymbol{x}}_{f_n}$ |
| $\boldsymbol{P}_{f_n}$ | covariance matrix of $\hat{\boldsymbol{x}}_{f_n}$ |
| $\boldsymbol{P}_{ft_n}, \boldsymbol{P}_{ftb_n}$ | submatrices of $\boldsymbol{P}_{f_n}$ |
| $\overline{\boldsymbol{P}}^*, \overline{\boldsymbol{P}}_n, \overline{\boldsymbol{P}}_{fb}$ | one-step prediction covariance matrices of $\boldsymbol{P}^*, \boldsymbol{P}_n, \boldsymbol{P}_{fb}$ |
| $[\overline{\boldsymbol{P}}^{*-1}]_{t_n},[\overline{\boldsymbol{P}}^{*-1}]_{t_{mn}},[\overline{\boldsymbol{P}}^{*-1}]_{tb_n},[\overline{\boldsymbol{P}}^{*-1}]_b$ | submatrices of $\overline{\boldsymbol{P}}^{*-1}$ |
| $[\overline{\boldsymbol{P}}_n^{-1}]_t,[\overline{\boldsymbol{P}}_n^{-1}]_{tb},[\overline{\boldsymbol{P}}_n^{-1}]_b$ | submatrices of $\overline{\boldsymbol{P}}_n^{-1}$ |
| $\boldsymbol{I}$ | identity matrix |
| $\boldsymbol{O}$ | all-zero matrix |
| $\boldsymbol{0}$ | all-zero vector |

### B. System Model

We start with linear state and measurement models. The case of nonlinear models will be discussed in Section IV. Suppose that there are $N$ targets. For the $n$th target, the state vector at time $k$ is denoted as $\boldsymbol{x}_{t_n}(k)$. The state equation of the $n$th target can be expressed as

$$\boldsymbol{x}_{t_n}(k+1) = \boldsymbol{F}_{t_n}(k)\boldsymbol{x}_{t_n}(k) + \boldsymbol{v}_{t_n}(k). \quad (1)$$

$\boldsymbol{F}_{t_n}(k)$ is the state transition matrix. The process noise $\boldsymbol{v}_{t_n}(k)$ is assumed to be a zero-mean white Gaussian noise with covariance $\boldsymbol{Q}_{t_n}(k)$, namely $\boldsymbol{v}_{t_n}(k) \sim \mathcal{N}(\boldsymbol{0}, \boldsymbol{Q}_{t_n}(k))$ and $E[\boldsymbol{v}_{t_n}(k)\boldsymbol{v}_{t_n}(l)^T] = \boldsymbol{Q}_{t_n}(k)\delta_{kl}$ with $\delta_{kl} = \begin{cases} 1, k=l \\ 0, k \neq l \end{cases}$. $\boldsymbol{0}$ is the all-zero column vector with corresponding size. $E[\cdot]$ denotes the statistical expectation, and superscript "T" denotes the transpose. Here the process noises of different targets are assumed to be statistically independent, namely $E[\boldsymbol{v}_{t_n}(k)\boldsymbol{v}_{t_m}(k)^T] = \boldsymbol{Q}_{t_n}(k)\delta_{nm}$.

Assume that the sensor biases are constant but unknown and only affect the measurements in the form of summation. Then the bias vector $\boldsymbol{b}$ conforms to

$$\boldsymbol{b}(k+1) = \boldsymbol{b}(k) = \boldsymbol{b}. \quad (2)$$

The measurement model of the multisensor system with respect to the $n$th target is expressed as

$$\boldsymbol{z}_n(k) = \boldsymbol{H}_{t_n}(k)\boldsymbol{x}_{t_n}(k) + \boldsymbol{H}_{b_n}(k)\boldsymbol{b} + \boldsymbol{w}_n(k). \quad (3)$$

For the multisensor case, $z_n(k)$ stacks the measurements of all the sensors with respect to the *n*th target. $H_{t_n}(k)$, $H_{b_n}(k)$, and $w_n(k)$ denotes the state measurement matrix, bias measurement matrix, and measurement noise, respectively. It is assumed that $w_n(k) \sim \mathcal{N}(0, R_n(k))$, $E[w_n(k)w_m(k)^T] = R_n(k)\delta_{nm}$, $E[w_n(k)w_n(l)^T] = R_n(k)\delta_{kl}$, and $E[w_n(k+1)v_{t_m}(l)^T] = O$. $O$ is the all-zero matrix with corresponding size.

Besides, it is assumed that the counterpart deterministic system of (1)-(3), namely the counterpart system without the process noise and measurement noise, is observable [1]. It indicates that the observability matrix composed of $H_{t_n}(k)$, $H_{b_n}(k)$, and $F_{t_n}(k)$, $k = 1, 2, \ldots$, has full column rank [25]. Under the observability assumption, the state estimator, e.g. the KF, can then be applied. In the following text, for description brevity, the time index in the $F_{t_n}(k)$, $v_{t_n}(k)$, $Q_{t_n}(k)$, $H_{t_n}(k)$, $H_{b_n}(k)$, $w_n(k)$, and $R_n(k)$ will be omitted in the case of no ambiguity.

### C. Augmented State Kalman Filter

In the ASKF, the augmented state stacks the states of all the targets and the sensor biases into a single vector, denoted by $x = [x_{t_1}^T, \ldots, x_{t_N}^T, b^T]^T$. The state equation of the augmented state $x$ is expressed as

$$x(k+1) = F(k)x(k) + v(k). \quad (4)$$

$F = \text{blkdiag}\{F_{t_1}, \cdots, F_{t_N}, I\}$, where $\text{blkdiag}\{\cdot\}$ denotes block diagonal matrix constructed by the matrices in the brace and $I$ is the identity matrix with appropriate size. $v = [v_{t_1}^T, \ldots, v_{t_N}^T, 0^T]^T$. The covariance matrix of $v$ becomes $Q = \text{blkdiag}\{Q_{t_1}, \ldots, Q_{t_N}, O\}$.

Likewise, stacking the measurements of all the targets into a single vector, denoted by $z = [z_1^T, \ldots, z_N^T]^T$, the augmented measurement equation can be expressed as

$$z(k) = H(k)x(k) + w(k), \quad (5)$$

where $H$ conforms to

$$H = \begin{bmatrix} H_{t_1} & \cdots & O & H_{b_1} \\ \vdots & \ddots & \vdots & \vdots \\ O & \cdots & H_{t_N} & H_{b_N} \end{bmatrix}. \quad (6)$$

$w = [w_1^T, \ldots, w_N^T]^T$. The covariance matrix of $w$ becomes $R = \text{blkdiag}\{R_1, \ldots, R_N\}$.

Given the augmented state and measurement models in (4) and (5), the ASKF can be expressed as follows

$$\hat{x}^*(k+1|k) = F(k)\hat{x}^*(k|k), \quad (7)$$

$$P^*(k+1|k) = F(k)P^*(k|k)F(k)^T + Q(k), \quad (8)$$

$$K(k+1) = P^*(k+1|k)H(k+1)^T \\ \times [H(k+1)P^*(k+1|k)H(k+1)^T + R(k+1)]^{-1}, \quad (9)$$

$$\hat{x}^*(k+1|k+1) = \hat{x}^*(k+1|k) \\ + K(k+1)[z(k+1) - H(k+1)\hat{x}^*(k+1|k)], \quad (10)$$

$$P^*(k+1|k+1) = [I - K(k+1)H(k+1)]P^*(k+1|k). \quad (11)$$

Note that superscript "*" has used to denote the estimate and covariance matrix of the ASKF. Besides, there is an equivalent information form of (10) and (11), i.e.

$$P^*(k+1|k+1)^{-1}\hat{x}^*(k+1|k+1) = \\ P^*(k+1|k)^{-1}\hat{x}^*(k+1|k) + H(k+1)^T R(k+1)^{-1} z(k+1), \quad (12)$$

$$P^*(k+1|k+1)^{-1} = P^*(k+1|k)^{-1} \\ + H(k+1)^T R(k+1)^{-1} H(k+1). \quad (13)$$

This information form of the KF is used hereafter to facilitate the following proofs.

### III. PROPOSED DECOUPLED KALMAN FILTERING

#### A. Algorithm Description

Instead of processing the multitarget states and sensor biases together as in the ASKF, a decoupled KF that deals with each target separately is proposed herein. A single-target KF branch is assigned for each target. The block diagram of the decoupled KF is presented in Fig. 1. There are three main modules, namely the augmented state estimation, bias information fusion, and augmented state update. First, the augmented state estimation is conducted for each single-target KF branch. The bias vector estimates of all the single-target KF branches are fused in the module of bias information fusion. The fused bias vector estimate is then fed back to each single-target KF branch to refine the state estimation.

The augmented state herein is different from that of the ASKF in Section II-C. It is only constructed by the states of a single target and the sensor biases. Let $x_n = [x_{t_n}^T, b^T]^T$ be the augmented state of the *n*th single-target KF branch. The corresponding state equation is expressed as

$$x_n(k+1) = F_n(k)x_n(k) + v_n(k), \quad (14)$$

where $F_n = \text{blkdiag}\{F_t, I\}$, and $v_n = [v_{t_n}^T, 0^T]^T$. The covariance of $v_n$ becomes $Q_n = \text{blkdiag}\{Q_{t_n}, O\}$.

The corresponding measurement equation is given in (3) and rewritten here for convenience, i.e.

$$z_n(k) = H_n(k)x_n(k) + w_n(k), \quad (15)$$

where $H_n = [H_{t_n}, H_{b_n}]$.

Let $\hat{x}_{f_n}(k|k)$ be the updated state of the *n*th single-target KF branch after the module of augmented state update and $P_{f_n}(k|k)$ be the covariance matrix of $\hat{x}_{f_n}(k|k)$.

1) Augmented state estimation

Applying the KF to model (14) and (15), the augmented state estimation in each single-target KF branch is expressed as

$$\hat{x}_n(k+1|k) = F_n(k)\hat{x}_{f_n}(k|k), \quad (16)$$

$$P_n(k+1|k) = F_n(k)P_{f_n}(k|k)F_n(k)^T + Q_n(k), \quad (17)$$

---

[1] Strictly speaking, the models (1)-(3) constitute a linear stochastic system. Regarding the stochastic observability, there are various definitions. One can referring to [26]-[29] and the references therein.

$$K_n(k+1) = P_n(k+1|k)H_n(k+1)^T$$
$$\times \left[ H_n(k+1)P_n(k+1|k)H_n(k+1)^T + R_n(k+1) \right]^{-1}, \quad (18)$$

$$\hat{x}_n(k+1|k+1) = \hat{x}_n(k+1|k) + K_n(k+1)\left[ z_n(k+1) - H_n(k+1)\hat{x}_n(k+1|k) \right], \quad (19)$$

$$P_n(k+1|k+1) = \left[ I - K_n(k+1)H_n(k+1) \right] P_n(k+1|k). \quad (20)$$

Likewise, the information form of (19) and (20) is

$$P_n(k+1|k+1)^{-1}\hat{x}_n(k+1|k+1) = P_n(k+1|k)^{-1}\hat{x}_n(k+1|k) + H_n(k+1)^T R_n(k+1)^{-1} z_n(k+1), \quad (21)$$

$$P_n(k+1|k+1)^{-1} = P_n(k+1|k)^{-1} + H_n(k+1)^T R_n(k+1)^{-1} H_n(k+1). \quad (22)$$

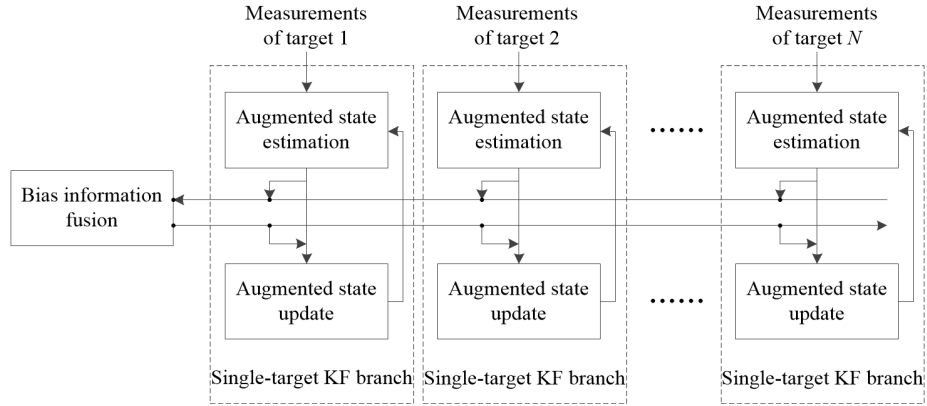

Fig. 1. The block diagram of the decoupled KF.

2) Bias information fusion

Partitioning $\hat{x}_n$ and $P_n$ according to the target states and sensor biases, we have $\hat{x}_n \triangleq \begin{bmatrix} \hat{x}_{t_n} \\ \hat{b}_n \end{bmatrix}$ and $P_n = \begin{bmatrix} P_{t_n} & P_{tb_n} \\ P_{tb_n}^T & P_{b_n} \end{bmatrix}$. For $N$ targets, there are $N$ separate estimates of the sensor bias vector, namely $\hat{b}_n$, $n=1,\ldots,N$. In this module, we fuse all the separate bias vector estimates. Let $\hat{b}_f$ be the fused bias vector estimate and $P_{fb}$ be the covariance matrix of $\hat{b}_f$. The bias information fusion is implemented as follows

$$\hat{b}_f(k+1|k) = \hat{b}_f(k|k), \quad (23)$$
$$P_{fb}(k+1|k) = P_{fb}(k|k), \quad (24)$$

$$\hat{b}_f(k+1|k+1) = P_{fb}(k+1|k+1)\Big\{ P_{fb}(k+1|k)^{-1}\hat{b}_f(k+1|k) + \sum_{n=1}^{N}\Big[ P_{b_n}(k+1|k+1)^{-1}\hat{b}_n(k+1|k+1) - P_{b_n}(k+1|k)^{-1}\hat{b}_n(k+1|k) \Big] \Big\}, \quad (25)$$

$$P_{fb}(k+1|k+1)^{-1} = P_{fb}(k+1|k)^{-1} + \sum_{n=1}^{N}\Big[ P_{b_n}(k+1|k+1)^{-1} - P_{b_n}(k+1|k)^{-1} \Big]. \quad (26)$$

3) Augmented state update

In this module, the augmented state of the each single-target KF branch is updated by applying the fused bias vector estimate. It is implemented as follows

$$\hat{x}_{f_n}(k+1|k+1) = \hat{x}_n + \begin{bmatrix} P_{tb_n} P_{b_n}^{-1}(\hat{b}_f - \hat{b}_n) \\ \hat{b}_f - \hat{b}_n \end{bmatrix} = \begin{bmatrix} \hat{x}_{t_n} + P_{tb_n} P_{b_n}^{-1}(\hat{b}_f - \hat{b}_n) \\ \hat{b}_f \end{bmatrix} \triangleq \begin{bmatrix} \hat{x}_{ft_n} \\ \hat{b}_f \end{bmatrix}, \quad (27)$$

$$P_{f_n}(k+1|k+1) = P_n - \begin{bmatrix} P_{tb_n} P_{b_n}^{-1}(P_{b_n} - P_{fb}) P_{b_n}^{-1} P_{tb_n}^T & P_{tb_n} P_{b_n}^{-1}(P_{b_n} - P_{fb}) \\ (P_{b_n} - P_{fb}) P_{b_n}^{-1} P_{tb_n}^T & P_{b_n} - P_{fb} \end{bmatrix} = \begin{bmatrix} P_{t_n} - P_{tb_n} P_{b_n}^{-1}(P_{b_n} - P_{fb}) P_{b_n}^{-1} P_{tb_n}^T & P_{tb_n} P_{b_n}^{-1} P_{fb} \\ P_{fb} P_{b_n}^{-1} P_{tb_n}^T & P_{fb} \end{bmatrix} \triangleq \begin{bmatrix} P_{ft_n} & P_{ftb_n} \\ P_{ftb_n}^T & P_{fb} \end{bmatrix}, \quad (28)$$

where the time indices $(k+1|k+1)$ of all the quantities in the right side of equal signs in (27) and (28) have been omitted to simplify the notation.

Note that the updated bias vector estimate and its covariance matrix are equal to that of the fused bias vector estimate. In addition, applying the matrix inversion formula in block form (called block matrix inversion formula hereafter) [30], eq. (28) can be equivalently expressed in the form of information matrix, i.e.

$$P_{fn}^{-1} = P_n^{-1} + \begin{bmatrix} O & O \\ O & P_{fb}^{-1} - P_{b_n}^{-1} \end{bmatrix}. \quad (29)$$

The derivation of (29) is detailed in the Appendix A. Note that only the bias submatrix of the information matrix changes after the augmented state update. It is consistent with the intuition that only the bias information is fed back for the update.

B. *Computational Complexity*

In the proposed decoupled KF, only the bias vector estimates and covariance matrices are needed to be transferred between the single-target KF branches and the bias information fusion module. This structure is very clear and flexible. Moreover, in this subsection, we will show that the proposed algorithm has lower computational complexity than the ASKF.

The number of multiplications required for one data update is adopted as a measure of computational complexity. Let $S$, $B$, and $M$ be the dimensions of $x_{t_n}$, $b$, and $z_n$, respectively. Table I and II summarize the computational complexity of the ASKF and proposed decoupled KF, respectively. The computational complexity of the proposed decoupled KF is linear with the number of targets, and it is two orders lower than that of the ASKF.

TABLE I
Computational complexity of the ASKF in term of multiplications required for one data update

| Processing Steps | Multiplications |
|---|---|
| $\hat{x}^*(k+1\|k) = F\hat{x}^*(k\|k)$ | $C_1 = NS^2$ |
| $\bar{P}^* = FP^*(k\|k)F^T + Q$ | $C_2 = 2NS^2(NS+B)$ |
| $K = \bar{P}^*H^T\left[H\bar{P}^*H^T + R\right]^{-1}$ | $C_3 = O(M^3N^3) + M^2N^2(NS+B)$ $+ 2MN^2S^2 + M^2N^2S$ $+ MN^2SB + 2M^2N^2B$ $+ 2MNSB + 2MNB^2$ |
| $\hat{x}^* = \hat{x}^*(k+1\|k)$ $+ K\left[z - H\hat{x}^*(k+1\|k)\right]$ | $C_4 = MN(NS+S+2B)$ |
| $P^* = (I - KH)P^*(k+1\|k)$ | $C_5 = MN(NS+B)^2 + MN^2S^2$ $+ MN^2SB + MNSB + MNB^2$ |
| Total Complexity | $C = C_1 + C_2 + C_3 + C_4 + C_5$ |
| Items of the highest order | $O(M^3N^3) + M^2N^2(NS+B) + MN(NS+B)^2$ |

Note: For fair comparison, the special structure of $F$ and $H$ has been considered in the statistics. $O(M^3N^3)$ denotes the computational complexity of the inversion of a matrix with size $MN \times MN$.

TABLE II
Computational complexity of the proposed decoupled KF in term of multiplications required for one data update

| Processing Steps | Multiplications |
|---|---|
| Augmented state estimation for a single-target KF branch | $C_1' = O(M^3) + 2M^2(S+B) + 3M(S+B)^2$ $+ 3(S+B)^3 + 2M(S+B) + (S+B)^2$ |
| Repeat $N$ times | $C_2' = NC_1'$ |
| Bias information fusion | $C_3' = (2N+2)O(B^3) + (2N+2)B^2$ |
| Augmented state update for a single-target KF branch | $C_4' = S^2B + 3SB^2 + SB$ |
| Repeat $N$ times | $C_5' = NC_4'$ |
| Total Complexity | $C' = C_2' + C_3' + C_5'$ |
| Items of the highest order | $N \times [O(M^3) + 2M^2(S+B) + 3M(S+B)^2$ $+ 3(S+B)^3 + S^2B + 3SB^2] + (2N+2)O(B^3)$ |

## IV. PROOF OF EQUIVALENCE BETWEEN PROPOSED DECOUPLED KF AND ASKF

### A. Proof of Equivalence Between Proposed Decoupled KF and ASKF

Let the initial state of the ASKF be
$$\hat{x}^*(0) = [\hat{x}_{t_1}^T(0), \cdots, \hat{x}_{t_N}^T(0), \hat{b}^T(0)]^T, \quad (30)$$

and the corresponding covariance matrix be
$$P^*(0) = \begin{bmatrix} \ddots & \cdots & \cdots & \cdots & \cdots & \vdots \\ \vdots & P_{t_m}(0) & \ddots & P_{t_{mn}}(0) & \ddots & P_{tb_m}(0) \\ \vdots & \ddots & \ddots & \ddots & \ddots & \vdots \\ \vdots & P_{t_{mn}}^T(0) & \ddots & P_{t_n}(0) & \ddots & P_{tb_n}(0) \\ \vdots & \ddots & \ddots & \ddots & \ddots & \vdots \\ \cdots & P_{tb_m}^T(0) & \cdots & P_{tb_n}^T(0) & \cdots & P_b(0) \end{bmatrix}, \quad (31)$$

where $P_{t_{mn}}$ denotes the cross-covariance matrix between the $m$th and $n$th targets' states, and $P_{tb_n}$ denotes the cross-covariance matrix between the $n$th target's states and sensor biases. We use $P_b(0)$ to represent the initial uncertainty of the sensor biases. Under the framework of KF, the sensor biases are also treated as Gaussian random variables although they are assumed to be unknown constant in the model.

Likewise, let the initial states of the decoupled KF be
$$\hat{x}_{f_n}(0) = [\hat{x}_{t_n}^T(0), \hat{b}^T(0)]^T, n = 1, \cdots, N, \quad (32)$$
$$\hat{b}_f(0) = \hat{b}(0), \quad (33)$$
and the corresponding covariance matrices be
$$P_{f_n}(0) = \begin{bmatrix} P_{t_n}(0) & P_{tb_n}(0) \\ P_{tb_n}^T(0) & P_b(0) \end{bmatrix}, n = 1, \cdots, N, \quad (34)$$
$$P_{fb}(0) = P_b(0). \quad (35)$$

Note that it is natural to set the same initial states and covariance matrix for the ASKF and the decoupled KF. To prove the equivalence between the proposed decoupled KF and the ASKF, we first prove the following lemma. In the following lemma, theorem, and proofs, to simplify the notation, the time index $(k+1|k+1)$ is omitted in the case of no ambiguity, e.g. $P^* \triangleq P^*(k+1|k+1)$, and the time index $(k+1|k)$ in the covariance matrix is replaced by a transverse line on the top of the symbol, e.g. $\bar{P}^* \triangleq P^*(k+1|k)$.

*Lemma*: Under the model (1)-(3) and the initialization (30)-(35), if $P_{t_{mn}}(0) = P_{tb_m}(0)P_b(0)^{-1}P_{tb_n}^T(0)$, $\forall m,n \in \{1,\ldots,N\}$, then $P_{t_{mn}}^*(k|k) = P_{tb_m}^*(k|k)P_b^*(k|k)^{-1}P_{tb_n}^{*T}(k|k)$, $\forall m,n \in \{1,\ldots,N\}, k \geq 0$.

Proof: We propose an inductive proof. According to the condition of the lemma, we obtain that $P_{t_{mn}}^*(k|k) = P_{tb_m}^*(k|k)P_b^*(k|k)^{-1}P_{tb_n}^{*T}(k|k)$, $\forall m,n$ holds when $k = 0$. Thus, we only need to prove that $P_{t_{mn}}^* = P_{tb_m}^* P_b^{*-1} P_{tb_n}^{*T}$, $\forall m,n$ for the time index $(k+1|k+1)$ also holds if $P_{t_{mn}}^*(k|k) = P_{tb_m}^*(k|k)P_b^*(k|k)^{-1}P_{tb_n}^{*T}(k|k)$, $\forall m,n$.

Without loss of generality, we first take the case of 2 targets for the proof. In this case, we partition all the covariance matrix according to the states of each target and the sensor biases. For example, $P^*(k|k)$, $\bar{P}^*$, $\bar{P}^{*-1}$, and $P^*$ can be expressed in the block matrix form

$$P^*(k|k) = \begin{bmatrix} P^*_{t_1}(k|k) & P^*_{t_{12}}(k|k) & P^*_{tb_1}(k|k) \\ P^{*T}_{t_{12}}(k|k) & P^*_{t_2}(k|k) & P^*_{tb_2}(k|k) \\ P^{*T}_{tb_1}(k|k) & P^{*T}_{tb_2}(k|k) & P^*_b(k|k) \end{bmatrix}, \quad (36)$$

$$\bar{P}^* = \begin{bmatrix} \bar{P}^*_{t_1} & \bar{P}^*_{t_{12}} & \bar{P}^*_{tb_1} \\ \bar{P}^{*T}_{t_{12}} & \bar{P}^*_{t_2} & \bar{P}^*_{tb_2} \\ \bar{P}^{*T}_{tb_1} & \bar{P}^{*T}_{tb_2} & \bar{P}^*_b \end{bmatrix}, \quad (37)$$

$$\bar{P}^{*-1} = \begin{bmatrix} [\bar{P}^{*-1}]_{t_1} & [\bar{P}^{*-1}]_{t_{12}} & [\bar{P}^{*-1}]_{tb_1} \\ [\bar{P}^{*-1}]^T_{t_{12}} & [\bar{P}^{*-1}]_{t_2} & [\bar{P}^{*-1}]_{tb_2} \\ [\bar{P}^{*-1}]^T_{tb_1} & [\bar{P}^{*-1}]^T_{tb_2} & [\bar{P}^{*-1}]_b \end{bmatrix}, \quad (38)$$

$$P^* = \begin{bmatrix} P^*_{t_1} & P^*_{t_{12}} & P^*_{tb_1} \\ P^{*T}_{t_{12}} & P^*_{t_2} & P^*_{tb_2} \\ P^{*T}_{tb_1} & P^{*T}_{tb_2} & P^*_b \end{bmatrix}, \quad (39)$$

where submatrices with subscripts "$t_1$", "$t_2$", and "$b$" denote the covariance submatrices corresponding to target 1, target 2, and bias vector, respectively. The submatrix with subscript "$t_{12}$" denotes the cross-covariance submatrix between the states of target 1 and target 2. The submatrix with subscript "$tb_n$" ($n = 1, 2$) denotes the cross-covariance submatrix between the states of target $n$ and the bias vector. $\bar{P}^{*-1}$ represents the corresponding information matrix of $\bar{P}^*$.

Substituting (36) into (8), $\bar{P}^* \triangleq P^*(k+1|k)$ conforms to

$$\bar{P}^* = \begin{bmatrix} F_{t_1} P^*_{t_1}(k|k) F^T_{t_1} + Q_{t_1} & F_{t_1} P^*_{t_{12}}(k|k) F^T_{t_2} & F_{t_1} P^*_{tb_1}(k|k) \\ F_{t_2} P^{*T}_{t_{12}}(k|k) F^T_{t_1} & F_{t_2} P^*_{t_2}(k|k) F^T_{t_2} + Q_{t_2} & F_{t_2} P^*_{tb_2}(k|k) \\ P^{*T}_{tb_1}(k|k) F^T_{t_1} & P^{*T}_{tb_2}(k|k) F^T_{t_2} & P^*_b(k|k) \end{bmatrix}. \quad (40)$$

According to $P^*_{t_{12}}(k|k) = P^*_{tb_1}(k|k) P^*_b(k|k)^{-1} P^{*T}_{tb_2}(k|k)$, it can be readily verified in $\bar{P}^*$ that

$$\bar{P}^*_{t_{12}} = \bar{P}^*_{tb_1} \bar{P}^{*-1}_b \bar{P}^{*T}_{tb_2}. \quad (41)$$

Applying the block matrix inversion formula to (37) and making it correspond to (38), we obtain

$$\begin{bmatrix} [\bar{P}^{*-1}]_{t_1} & [\bar{P}^{*-1}]_{t_{12}} \\ [\bar{P}^{*-1}]^T_{t_{12}} & [\bar{P}^{*-1}]_{t_2} \end{bmatrix} = \left\{ \begin{bmatrix} \bar{P}^*_{t_1} & \bar{P}^*_{t_{12}} \\ \bar{P}^{*T}_{t_{12}} & \bar{P}^*_{t_2} \end{bmatrix} - \begin{bmatrix} \bar{P}^*_{tb_1} \\ \bar{P}^*_{tb_2} \end{bmatrix} \bar{P}^{*-1}_b \begin{bmatrix} \bar{P}^{*T}_{tb_1} & \bar{P}^{*T}_{tb_2} \end{bmatrix} \right\}^{-1} = \begin{bmatrix} \left(\bar{P}^*_{t_1} - \bar{P}^*_{tb_1} \bar{P}^{*-1}_b \bar{P}^{*T}_{tb_1}\right)^{-1} & O \\ O & \left(\bar{P}^*_{t_2} - \bar{P}^*_{tb_2} \bar{P}^{*-1}_b \bar{P}^{*T}_{tb_2}\right)^{-1} \end{bmatrix}, \quad (42)$$

$$\begin{bmatrix} [\bar{P}^{*-1}]_{tb_1} \\ [\bar{P}^{*-1}]_{tb_2} \end{bmatrix} = \begin{bmatrix} -\left(\bar{P}^*_{t_1} - \bar{P}^*_{tb_1} \bar{P}^{*-1}_b \bar{P}^{*T}_{tb_1}\right)^{-1} \bar{P}^*_{tb_1} \bar{P}^{*-1}_b \\ -\left(\bar{P}^*_{t_2} - \bar{P}^*_{tb_2} \bar{P}^{*-1}_b \bar{P}^{*T}_{tb_2}\right)^{-1} \bar{P}^*_{tb_2} \bar{P}^{*-1}_b \end{bmatrix}, \quad (43)$$

$$[\bar{P}^{*-1}]_b = \bar{P}^{*-1}_b + \bar{P}^{*-1}_b \begin{bmatrix} \bar{P}^{*T}_{tb_1} & \bar{P}^{*T}_{tb_2} \end{bmatrix} \left\{ \begin{bmatrix} \bar{P}^*_{t_1} & \bar{P}^*_{t_{12}} \\ \bar{P}^{*T}_{t_{12}} & \bar{P}^*_{t_2} \end{bmatrix} - \begin{bmatrix} \bar{P}^*_{tb_1} \\ \bar{P}^*_{tb_2} \end{bmatrix} \bar{P}^{*-1}_b \begin{bmatrix} \bar{P}^{*T}_{tb_1} & \bar{P}^{*T}_{tb_2} \end{bmatrix} \right\}^{-1} \begin{bmatrix} \bar{P}^*_{tb_1} \\ \bar{P}^*_{tb_2} \end{bmatrix} \bar{P}^{*-1}_b$$
$$= \bar{P}^{*-1}_b + \bar{P}^{*-1}_b \bar{P}^{*T}_{tb_1} \left(\bar{P}^*_{t_1} - \bar{P}^*_{tb_1} \bar{P}^{*-1}_b \bar{P}^{*T}_{tb_1}\right)^{-1} \bar{P}^*_{tb_1} \bar{P}^{*-1}_b + \bar{P}^{*-1}_b \bar{P}^{*T}_{tb_2} \left(\bar{P}^*_{t_2} - \bar{P}^*_{tb_2} \bar{P}^{*-1}_b \bar{P}^{*T}_{tb_2}\right)^{-1} \bar{P}^*_{tb_2} \bar{P}^{*-1}_b, \quad (44)$$

where (41) has been applied in (42). According to (42), there is $[\bar{P}^{*-1}]_{t_{12}} = O$. $\bar{P}^{*-1}$ can be expressed as

$$\bar{P}^{*-1} = \begin{bmatrix} [\bar{P}^{*-1}]_{t_1} & O & [\bar{P}^{*-1}]_{tb_1} \\ O & [\bar{P}^{*-1}]_{t_2} & [\bar{P}^{*-1}]_{tb_2} \\ [\bar{P}^{*-1}]^T_{tb_1} & [\bar{P}^{*-1}]^T_{tb_2} & [\bar{P}^{*-1}]_b \end{bmatrix}. \quad (45)$$

Further substituting (45) into (13), we obtain

$$P^{*-1} = \bar{P}^{*-1} + H^T R^{-1} H = \begin{bmatrix} [\bar{P}^{*-1}]_{t_1} + H^T_{t_1} R^{-1}_1 H_{t_1} & O & [\bar{P}^{*-1}]_{tb_1} + H^T_{t_1} R^{-1}_1 H_{b_1} \\ O & [\bar{P}^{*-1}]_{t_2} + H^T_{t_2} R^{-1}_2 H_{t_2} & [\bar{P}^{*-1}]_{tb_2} + H^T_{t_2} R^{-1}_2 H_{b_2} \\ [\bar{P}^{*-1}]^T_{tb_1} + H^T_{b_1} R^{-1}_1 H_{t_1} & [\bar{P}^{*-1}]^T_{tb_2} + H^T_{b_2} R^{-1}_2 H_{t_2} & [\bar{P}^{*-1}]_b + H^T_{b_1} R^{-1}_1 H_{b_1} + H^T_{b_2} R^{-1}_2 H_{b_2} \end{bmatrix}. \quad (46)$$

Applying the block matrix inversion formula to (39) and making it correspond to (46), we obtain

$$\left\{ \begin{bmatrix} P^*_{t_1} & P^*_{t_{12}} \\ P^{*T}_{t_{12}} & P^*_{t_2} \end{bmatrix} - \begin{bmatrix} P^*_{tb_1} \\ P^*_{tb_2} \end{bmatrix} P^{*-1}_b \begin{bmatrix} P^{*T}_{tb_1} & P^{*T}_{tb_2} \end{bmatrix} \right\}^{-1} = \begin{bmatrix} [\bar{P}^{*-1}]_{t_1} + H^T_{t_1} R^{-1}_1 H_{t_1} & O \\ O & [\bar{P}^{*-1}]_{t_2} + H^T_{t_2} R^{-1}_2 H_{t_2} \end{bmatrix}, \quad (47)$$

$$-\begin{bmatrix}[\bar{P}^{*-1}]_{t_1} + H_{t_1}^T R_1^{-1} H_{t_1} & O \\ O & [\bar{P}^{*-1}]_{t_2} + H_{t_2}^T R_2^{-1} H_{t_2}\end{bmatrix}\begin{bmatrix}P_{tb_1}^* \\ P_{tb_2}^*\end{bmatrix}P_b^{*-1} = \begin{bmatrix}[\bar{P}^{*-1}]_{tb_1} + H_{t_1}^T R_1^{-1} H_{b_1} \\ [\bar{P}^{*-1}]_{tb_2} + H_{t_2}^T R_2^{-1} H_{b_2}\end{bmatrix}. \quad (48)$$

Expanding (47), we get

$$P_{t_n}^* - P_{tb_n}^* P_b^{*-1} P_{tb_n}^{*T} = \left([\bar{P}^{*-1}]_{t_n} + H_{t_n}^T R_n^{-1} H_{t_n}\right)^{-1}, n=1,2, \quad (49)$$

$$P_{t_{12}}^* - P_{tb_1}^* P_b^{*-1} P_{tb_2}^{*T} = O. \quad (50)$$

Eq. (50) is the desired result of the lemma.

In addition, it is easy to verify that the above derivations can be extended to the case of $N$ targets ($N \geq 2$). The general processes are as follows. First, according to $P_{t_{mn}}^*(k|k) = P_{tb_m}^*(k|k)P_b^*(k|k)^{-1}P_{tb_n}^{*T}(k|k)$, $\forall m,n$, we can obtain $\bar{P}_{t_{mn}}^* = \bar{P}_{tb_m}^* \bar{P}_b^{*-1} \bar{P}_{tb_n}^{*T}$, $\forall m,n$, as in (41). Then treating the states of all the target as a block and applying the block matrix inversion formula, we can get a block diagonal matrix for the target state information submatrix as in (42). With the block diagonal target state information submatrix, we can further obtain the desired result of the lemma by following the processes as in (46)-(50). In summary, the lemma is proved.

Now we prove the equivalence between the proposed decoupled KF and the ASKF. It is expressed as the following theorem.

**Theorem**: Under the model (1)-(3), the initialization (30)-(35), and the initial condition $P_{t_{mn}}(0) = P_{tb_m}(0)P_b(0)^{-1}P_{tb_n}^T(0)$, $\forall m,n \in \{1,\ldots,N\}$, the proposed decoupled KF is equivalent to the ASKF.

Proof: Appendix B gives the detailed proof of the theorem when $N=2$. It is an inductive proof. The core is to prove that $\hat{x}_{ft_n} = \hat{x}_{t_n}^*$, $\hat{b}_f = \hat{b}^*$, $P_{ft_n} = P_{t_n}^*$, $P_{ftb_n} = P_{tb_n}^*$, $P_{fb} = P_b^*$, $n=1,2$, holds if $\hat{x}_{ft_n}(k|k) = \hat{x}_{t_n}^*(k|k)$, $\hat{b}_f(k|k) = \hat{b}^*(k|k)$, $P_{ft_n}(k|k) = P_{t_n}^*(k|k)$, $P_{ftb_n}(k|k) = P_{tb_n}^*(k|k)$, $P_{fb}(k|k) = P_b^*(k|k)$, $n=1,2$. To this end, we first prove $P_{fb} = P_b^*$ by comparing $P_{fb}^{-1}$ with $P_b^{*-1}$. It involves a series of formula expansions and block matrix inversions. Then based on $P_{fb} = P_b^*$, we further prove $\hat{b}_f = \hat{b}^*$ and $\hat{x}_{ft_n} = \hat{x}_{t_n}^*$. Finally, we prove $P_{ft_n} = P_{t_n}^*$ and $P_{ftb_n} = P_{tb_n}^*$.

In addition, it is easy to extend the proof to the case of $N \geq 2$. That is, we can treat all the $N$ targets as a super target. Then applying the proof to the super target and the $(N+1)$th target, we can prove the theorem for the case of $N+1$ targets. Thus, the theorem is proved.

*B. Discussions*

In the proofs, note that there is an important feature in the information matrix $\bar{P}^{*-1}$ and $P^{*-1}$, namely $[\bar{P}^{*-1}]_{t_{mn}} = O = [P^{*-1}]_{t_{mn}}$. The zero cross-information matrix between targets lays the foundation for an exactly or equivalently decoupled implementation.

The initial condition $P_{t_{mn}}(0) = P_{tb_m}(0)P_b(0)^{-1}P_{tb_n}^T(0)$, $\forall m,n \in \{1,\ldots,N\}$, plays an important role in guaranteeing the zero cross-information matrix between targets. It should be highlighted that $P_{t_{mn}}(0) = P_{tb_m}(0)P_b(0)^{-1}P_{tb_n}^T(0)$ is usually satisfied. For example, in the initialization, both the measurement noise $w_n(0)$ and the uncertainty of sensor biases $\Delta b(0) = b(0) - \hat{b}(0)$ are unknown factors in the measurement. For the initial state estimation, an alternative approach may treat these two unknown factors as the composite measurement noise, namely $w_n(0) + H_{b_n}\Delta b(0)$. By linearization, the initial state estimation error of the target, denoted as $\Delta x_{t_n}(0) = x_{t_n}(0) - \hat{x}_{t_n}(0)$, can be expressed as $\Delta x_{t_n}(0) = G_n[w_n(0) + H_{b_n}\Delta b(0)]$, where $G_n$ denotes the gain matrix. Assuming that the uncertainty of sensor biases is independent of the measurement noise, we can obtain

$$P_{tb_n}(0) = G_n H_{b_n} P_b(0), \quad (51)$$

$$P_{t_{mn}}(0) = G_m H_{b_m} P_b(0) H_{b_n}^T G_n^T = P_{tb_m}(0) P_b(0)^{-1} P_{tb_n}^T(0), \quad (52)$$

where (51) has been applied in (52). Eq. (52) satisfies the required initial condition.

Another underlying condition to ensure the zero cross-information matrix between targets is the constant sensor bias model. It is the case for many applications. For the time-varying sensor biases, there is nonzero cross-information matrix between targets. The lemma would not exactly hold. The proposed decoupled KF becomes an approximation to the ASKF.

In addition, the above theorem and proof are based on the linear state and measurement models. For the nonlinear models, if extended Kalman filter (EKF) is applied to both the proposed decoupled KF and the ASKF, the equivalence between these two approaches can also be proved.

It is interesting to find that the bias fusion formulas (25) and (26) are quite similar to the fusion equations of the optimal distributed estimation in [31]. The application scenario of the optimal distributed estimation in [31] is that each sensor gets an estimate of the same target's states. The problem therein is to fuse these distributed state estimates to get a better state estimate. The problem discussed in this paper is different. The intuitive difference is that the bias vector estimates herein are separated according to targets, not sensors. Moreover, the bias vector estimates to be fused are only a part of the augmented state in each single-target KF branch. The problem herein is more complicated.

V. SIMULATION AND FIELD EXPERIMENTAL RESULTS

In this section, simulations are first conducted in the context of multistatic radars to verify the theoretical development and to compare the performance of the proposed algorithm with that of the ASKF and the approximately decoupled KF in [20]. The proposed algorithm is then applied to real data of a multistatic passive radar to test the practical feasibility.

### A. Simulations

The simulation scenario is shown in Fig. 2. Two-dimensional (2-D) space is considered in the simulation. There are 5 transmitters, 1 receiver, and 3 targets. Each transmitter-receiver (bistatic) pair composes a sensor. All the target trajectories continue 100 s. The data refresh period is 1 s. The target states are generated according to the Wiener-sequence acceleration model [32]. We enable all the bistatic pairs to detect all the targets to facilitate the execution of the ASKF. The data association is assumed to be known a priori.

In the simulation, the measurement consists of bistatic range and bistatic velocity. The bistatic range is the sum of the transmitter-to-target range and target-to-receiver range, namely $\|r - s_t\|_2 + \|r - s_r\|_2$, where $r$, $s_t$, and $s_r$ denote the locations of the target, transmitter, and receiver, respectively, and $\|\cdot\|_2$ denotes the Euclidean norm. The bistatic velocity is the rate of the bistatic range (related to the bistatic Doppler frequency). It can be expressed as $v^T[(s_t - r)/\|s_t - r\|_2 + (s_r - r)/\|s_r - r\|_2]$, where $v$ is the velocity of the target. Without loss of generality, only bistatic range biases are inserted. In each Monte-Carlo (MC) simulation, the bistatic range biases are generated according to zero-mean Gaussian distribution with standard deviation 300 m and are maintained constant over the 100 s. The measurement noises of the bistatic range and bistatic velocity follow the independent and identically distributed (IID) zero-mean Gaussian distribution with standard deviation 30 m and 1.5 m/s, respectively. 100 MC simulations are conducted.

In the simulation, the ASKF is used as the benchmark and the approximately decoupled KF in [20] is used for comparison. Note that the Kalman filters in all the methods are replaced by the EKF as the measurement is nonlinear with the target state.

The initialization follows (30)-(35). The initial bistatic range biases are set as zeros and the corresponding covariance matrix is set as an identity matrix multiplied (300 m)$^2$. Then the initial target states are estimated using the biased bistatic ranges and the bistatic velocities. The covariance matrix is calculated according to (51) and (52).

each time step using the 100 MC simulation results. Let $r_{t_n,m}$ be the location of target $n$ in the $m$th MC simulation. $\hat{r}_{t_n,m}$ denotes the estimate of $r_{t_n,m}$. The RMSE of target location estimation is expressed as

$$\text{RMSE}(\hat{r}_{t_n}) = \sqrt{\frac{1}{100}\sum_{m=1}^{100}\left\|\hat{r}_{t_n,m} - r_{t_n,m}\right\|^2}. \quad (53)$$

Likewise, let $\delta_m$ be the bistatic range bias vector in the $m$th MC simulation. $\hat{\delta}_m$ is the estimate of $\delta_m$. The RMSE of bistatic range bias estimation is expressed as

$$\text{RMSE}(\hat{\delta}) = \sqrt{\frac{1}{100}\sum_{m=1}^{100}\left\|\hat{\delta}_m - \delta_m\right\|^2 / 5}. \quad (54)$$

The other metric is the normalized estimation error squared (NEES) [33]. The NEES is used to check whether the estimator is consistent. The average NEES of target location estimation is defined as

$$\text{NEES}(\hat{r}_{t_n}) = \frac{1}{100}\sum_{m=1}^{100}(\hat{r}_{t_n,m} - r_{t_n,m})^T \text{Cov}(\hat{r}_{t_n,m})^{-1}(\hat{r}_{t_n,m} - r_{t_n,m}), \quad (55)$$

where $\text{Cov}(\hat{r}_{t_n,m})$ is the covariance matrix of $\hat{r}_{t_n,m}$ given by the estimator. Likewise, the average NEES of bistatic range bias estimation can also be expressed in the same way.

The RMSEs of the target location estimation and the bistatic range bias estimation are shown in Fig. 3 and Fig. 4, respectively. Note that only the result of target 1 is given to reduce the length as the results of the other two targets are similar. The RMSEs of the proposed decoupled KF algorithm exactly coincide with that of the ASKF, which is consistent with the theorem. Moreover, the RMSEs of the proposed algorithm are much lower than that of the approximately decoupled KF. Specifically, the RMSE of bistatic range bias of the proposed decoupled KF at 100 s is about 13.1 m while the one of the approximately decoupled KF is about 64.9 m. The proposed algorithm reduces the bistatic range errors about 80% compared with the approximately decoupled KF under the simulation scenario.

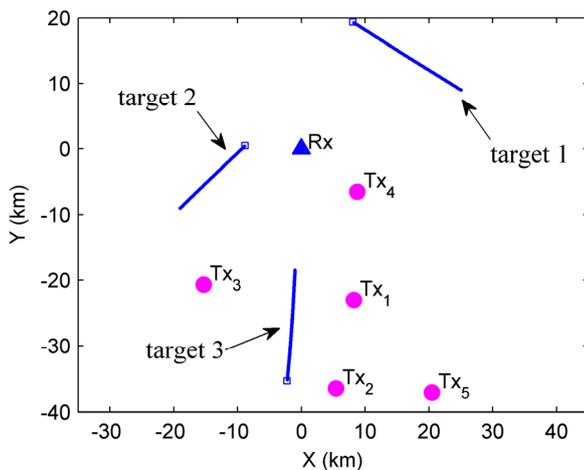

Fig. 2. The simulation scenario. A square is plotted at the terminal of each target trajectory.

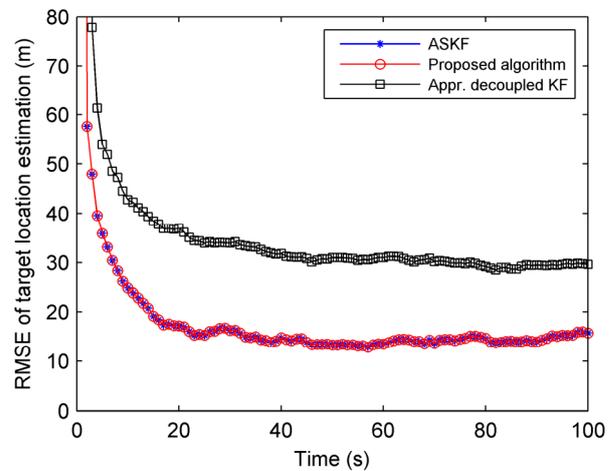

Fig. 3. The RMSEs of the target location estimation versus time of target 1.

Two metrics are utilized to measure the performance. One metric is the root mean square error (RMSE). It is calculated for

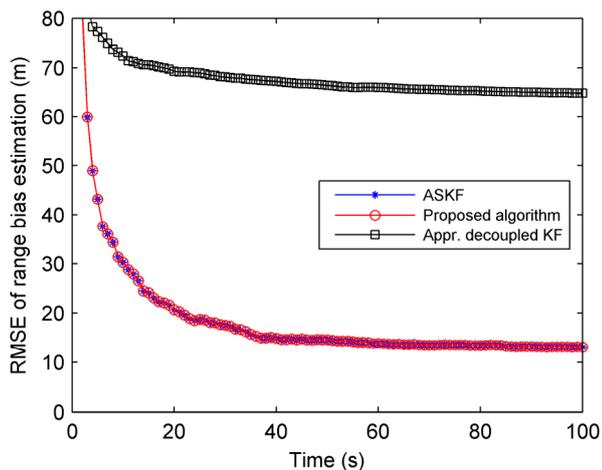

Fig. 4. The RMSEs of the bistatic range bias estimation versus time.

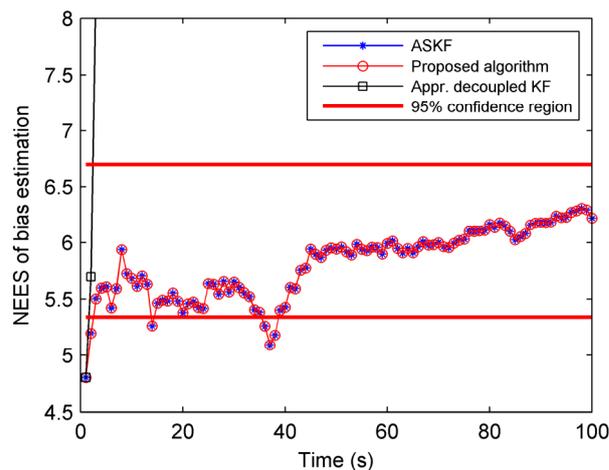

Fig. 6. The average NEESs of the bistatic range bias estimation versus time

The average NEESs of target location estimation and bistatic range bias estimation are shown in Fig. 5 and Fig. 6, respectively. The average NEESs of the proposed algorithm also exactly coincide with that of the ASKF. Besides, the two-side 95% confidence regions of the average NEESs are also marked by bold red lines. Almost all the average NEESs of the proposed algorithm lies in their 95% confidence region. It indicates that the proposed algorithm is consistent. By comparison, very few average NEESs of the approximately decoupled KF appear in the 95% confidence region, thereby inconsistent. It is caused by the ignored cross-correlation between target states and biases in the approximately decoupled KF.

In addition, it should be mentioned that the estimation results of the proposed algorithm are also the same with that of the ASKF when we compare each single MC simulation. It further confirms the theorem in Section IV.

### B. Field Experimental Results

We have developed an ultrahigh frequency (UHF) band passive radar system in Wuhan University. One set of the field experimental data in Nov. 2013 is utilized to test the proposed algorithm. The digital television network in Wuhan city, China, is exploited as the illuminators of opportunity in this experiment. The system configuration is with multiple transmitters and one receiver, as shown in Fig. 7. The signal bandwidth is about 7.5 MHz. Automatic dependent surveillance-broadcast (ADS-B) is utilized to record the civil airline information. There are 3 low-altitude aircrafts in this data. One of which has ADS-B reference information. The bistatic range bias references are calculated from the ADS-B reference information.

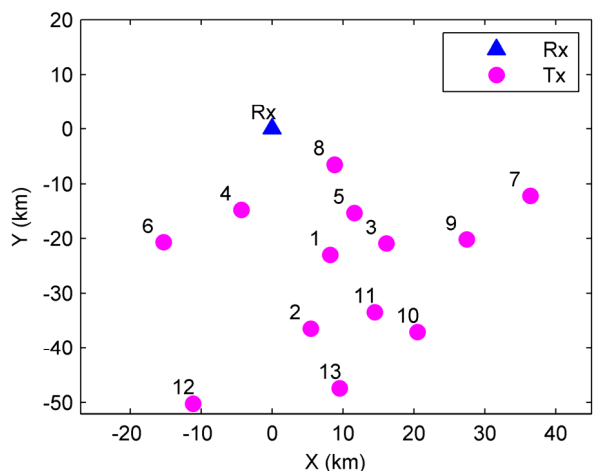

Fig. 7. The geometry of the field experiment.

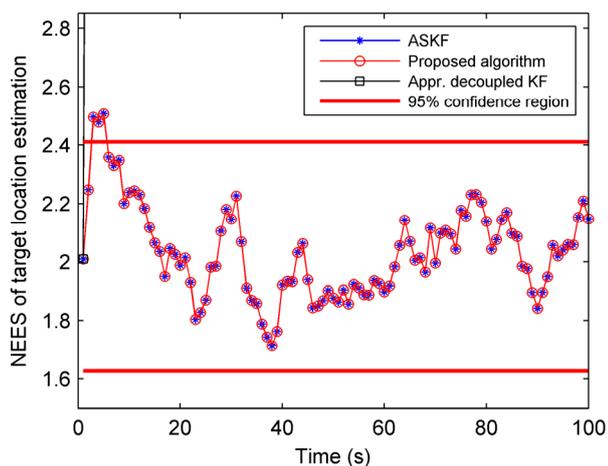

Fig. 5. The average NEESs of the target location estimation versus time of target 1.

In the real data, the detected targets are in low altitude due to the elevation coverage of the digital television transmitting antenna. Nevertheless, three-dimensional (3-D) space is still considered. We also use the bistatic ranges and bistatic velocities for the multi-target state estimation and bias estimation. As the bistatic velocities are almost unbiased due to the synchronization of the global positioning system (GPS), we only consider the bistatic range biases. The white noise jerk model [32] is used as the dynamic model for real-life targets. The initialization also follows (24)-(29). The initial bistatic range biases are set as zeros and the corresponding covariance matrix is set as identity matrix multiplied $(400 \text{ m})^2$. The measurement noise standard deviations of the bistatic range and bistatic velocity are set as 30 m and 1.5 m/s for the data processing.

Fig. 8 shows the bistatic range bias estimates and the corresponding error intervals of the proposed algorithm. For evaluation, the bias references obtained from the ADS-B information are also given. Most of the reference values lie within the corresponding error intervals of the estimates. It is clearer to draw this conclusion from Fig. 9 where the biases are aligned by shifting the reference values to zero. Specifically, the original bistatic range biases are distributed from −350 m to 150 m. After the estimation of the proposed algorithm, the residual bistatic range biases changes are distributed from −40 m to 5 m. It indicates that the bias estimation is effective.

The corresponding target tracks of the proposed algorithm are shown in Fig. 10. The black arrows point out the moving directions of the aircrafts. The ADS-B references of one of the landing aircraft are marked by red circles. Meanwhile, we also give the target tracks obtained without registration. For ease of observation, we give the zoomed figure at the ending part in Fig. 11. Many short and messy tracks can be observed in the case without registration. It is the typical influence of sensor biases. It originates from the variously biased state estimates at different frames. By contrast, the target tracks of the proposed algorithm are much better. The target tracks become continuous and almost coincide with the ADS-B references. Especially at the ending part, with the improvement of the bias estimation, the localization accuracy improves to about 50 m level. These results demonstrate that the proposed algorithm is effective for the real data.

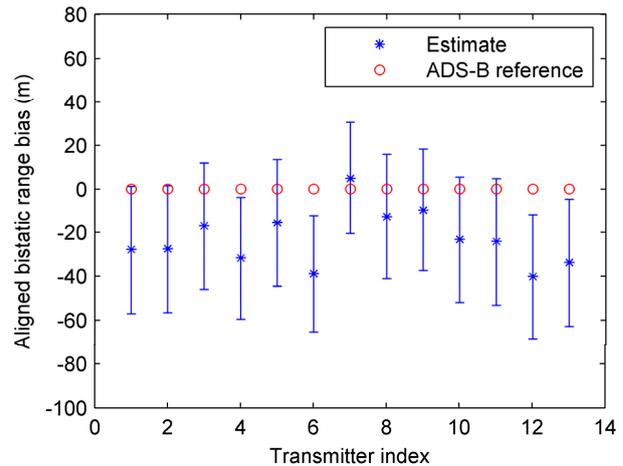
Fig. 9. Aligned bistatic range biases after shifting the reference values to zero.

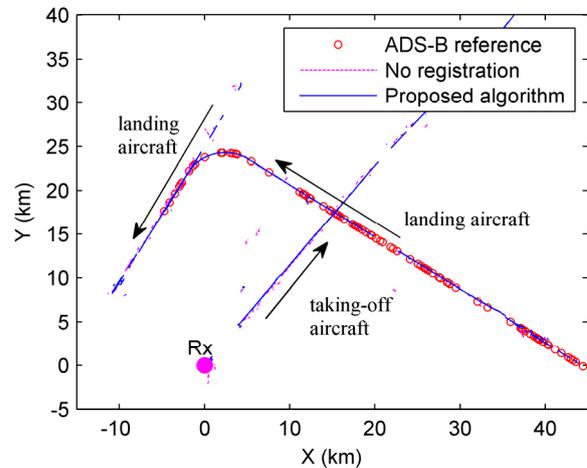
Fig. 10. Target tracks of the proposed algorithm and that without registration.

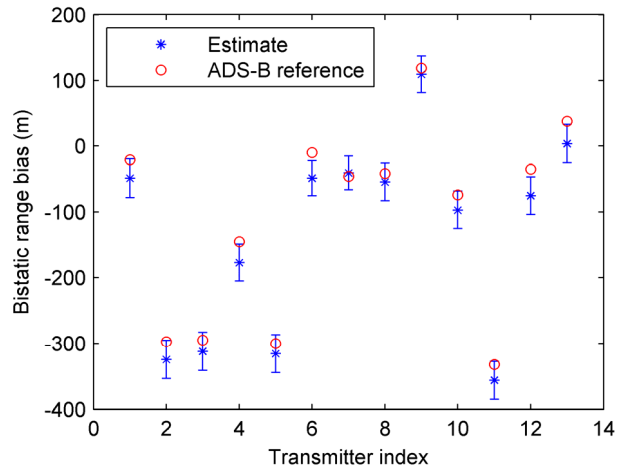
Fig. 8. The estimated bistatic range biases of real data.

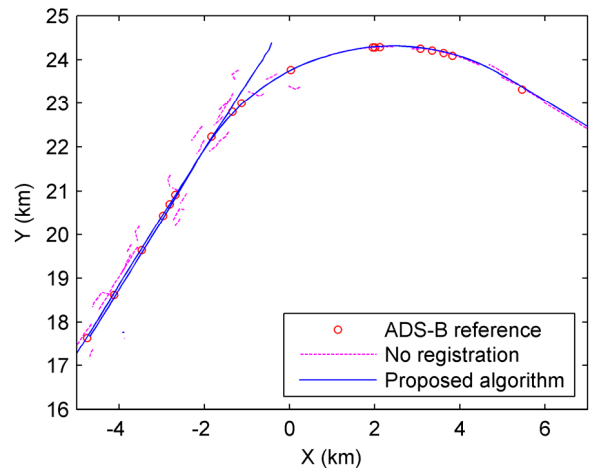
Fig. 11. Zoomed figure of Fig. 10 at the ending part.

## VI. CONCLUSIONS

In this paper we provide a novel decoupled KF for the multitarget state estimation in the presence of unknown sensor biases. This decoupled KF deals with each target separately except that the bias vector estimates of all the single-target KF branches are fused and the fused bias vector estimate is fed back to each single-target KF branch. It can readily handle the case with a dynamic change of the number of targets thanks to the decoupled structure. The computational complexity of the proposed decoupled KF is linear with the number of targets, and it is two orders lower than that of the ASKF. We prove that the proposed decoupled KF is exactly equivalent to the ASKF in terms of the estimation results under a usual initial condition. The simulation in the context of multistatic radars confirms the equivalence. It also shows that the proposed algorithm reduces the bistatic range bias errors about 80% compared with the approximately decoupled KF under the simulation scenario. More importantly, the proposed algorithm is also consistent with respect to the estimation error and corresponding covariance matrix. Field experimental data validates the practical feasibility of the proposed algorithm. In future work, further researches under the time-varying sensor bias model will be a useful extension. The joint state estimation, registration and data association is also an interesting topic.

## APPENDIX A. DERIVATION OF (29)

Assume that $P_{t_n}$ and $P_{b_n}$ are non-singular matrices. We first expand $P_n^{-1} = \begin{bmatrix} P_{t_n} & P_{tb_n} \\ P_{tb_n}^T & P_{b_n} \end{bmatrix}^{-1}$. The block matrix inversion formula is expressed in (56). Applying (56) to $P_n^{-1}$, we obtain (57). Then we analyze $P_{f_n}^{-1}$. Partition $P_{f_n}^{-1}$ according to the target states and biases, namely $P_{f_n}^{-1} = \begin{bmatrix} [P_{f_n}^{-1}]_t & [P_{f_n}^{-1}]_{tb} \\ [P_{f_n}^{-1}]_{tb}^T & [P_{f_n}^{-1}]_b \end{bmatrix}$. Likewise, applying (56) to (28), we obtain (58), (59), (60), and (61). Comparing (57) with (61), we obtain (29).

$$\begin{bmatrix} A & U \\ V & D \end{bmatrix}^{-1} = \begin{bmatrix} (A-UD^{-1}V)^{-1} & -(A-UD^{-1}V)^{-1}UD^{-1} \\ -D^{-1}V(A-UD^{-1}V)^{-1} & (D-VA^{-1}U)^{-1} \end{bmatrix} = \begin{bmatrix} (A-UD^{-1}V)^{-1} & -(A-UD^{-1}V)^{-1}UD^{-1} \\ -D^{-1}V(A-UD^{-1}V)^{-1} & D^{-1}+D^{-1}V(A-UD^{-1}V)^{-1}UD^{-1} \end{bmatrix}. \quad (56)$$

$$P_n^{-1} = \begin{bmatrix} (P_{t_n}-P_{tb_n}P_{b_n}^{-1}P_{tb_n}^T)^{-1} & -(P_{t_n}-P_{tb_n}P_{b_n}^{-1}P_{tb_n}^T)^{-1}P_{tb_n}P_{b_n}^{-1} \\ -P_{b_n}^{-1}P_{tb_n}^T(P_{t_n}-P_{tb_n}P_{b_n}^{-1}P_{tb_n}^T)^{-1} & P_{b_n}^{-1}+P_{b_n}^{-1}P_{tb_n}^T(P_{t_n}-P_{tb_n}P_{b_n}^{-1}P_{tb_n}^T)^{-1}P_{tb_n}P_{b_n}^{-1} \end{bmatrix}. \quad (57)$$

$$[P_{f_n}^{-1}]_t = \left[ P_{t_n} - P_{tb_n}P_{b_n}^{-1}(P_{b_n}-P_{fb})P_{b_n}^{-1}P_{tb_n}^T - P_{tb_n}P_{b_n}^{-1}P_{fb}P_{fb}^{-1}P_{fb}P_{b_n}^{-1}P_{tb_n}^T \right]^{-1} = (P_{t_n}-P_{tb_n}P_{b_n}^{-1}P_{tb_n}^T)^{-1}, \quad (58)$$

$$[P_{f_n}^{-1}]_{tb} = -[P_{f_n}^{-1}]_t P_{tb_n}P_{b_n}^{-1}P_{fb}P_{fb}^{-1} = -(P_{t_n}-P_{tb_n}P_{b_n}^{-1}P_{tb_n}^T)^{-1}P_{tb_n}P_{b_n}^{-1}, \quad (59)$$

$$[P_{f_n}^{-1}]_b = P_{fb}^{-1} + P_{fb}^{-1}P_{fb}P_{b_n}^{-1}P_{tb_n}^T U_{ft_n}P_{tb_n}P_{b_n}^{-1}P_{fb}P_{fb}^{-1} = P_{fb}^{-1} + P_{b_n}^{-1}P_{tb_n}^T(P_{t_n}-P_{tb_n}P_{b_n}^{-1}P_{tb_n}^T)^{-1}P_{tb_n}P_{b_n}^{-1}, \quad (60)$$

$$P_{f_n}^{-1} = \begin{bmatrix} (P_{t_n}-P_{tb_n}P_{b_n}^{-1}P_{tb_n}^T)^{-1} & -(P_{t_n}-P_{tb_n}P_{b_n}^{-1}P_{tb_n}^T)^{-1}P_{tb_n}P_{b_n}^{-1} \\ -P_{b_n}^{-1}P_{tb_n}^T(P_{t_n}-P_{tb_n}P_{b_n}^{-1}P_{tb_n}^T)^{-1} & P_{fb}^{-1}+P_{b_n}^{-1}P_{tb_n}^T(P_{t_n}-P_{tb_n}P_{b_n}^{-1}P_{tb_n}^T)^{-1}P_{tb_n}P_{b_n}^{-1} \end{bmatrix}. \quad (61)$$

## APPENDIX B. PROOF OF THE THEOREM WHEN $N = 2$

We propose an inductive proof for the case of 2 targets. According to the initialization (30)-(35), the initial states and covariance matrices of the decoupled KF is the same to that of the ASKF. Thus, we only need to prove that $\hat{x}_{ft_n} = \hat{x}_{t_n}^*$, $\hat{b}_f = \hat{b}^*$, $P_{ft_n} = P_{t_n}^*$, $P_{ftb_n} = P_{tb_n}^*$, $P_{fb} = P_b^*$, $n=1,2$ holds if
$\hat{x}_{ft_n}(k|k) = \hat{x}_{t_n}^*(k|k)$, $\hat{b}_f(k|k) = \hat{b}^*(k|k)$,
$P_{ft_n}(k|k) = P_{t_n}^*(k|k)$, $P_{ftb_n}(k|k) = P_{tb_n}^*(k|k)$,
$P_{fb}(k|k) = P_b^*(k|k)$, $n=1,2$.

According to the one-step prediction equations (7), (8), (16), (17), (23), and (24), we can obtain that $\hat{x}_{t_n}(k+1|k) = \hat{x}_{t_n}^*(k+1|k)$, $\hat{b}_f(k+1|k) = \hat{b}^*(k+1|k)$, $\overline{P}_{t_n} = \overline{P}_{t_n}^*$, $\overline{P}_{tb_n} = \overline{P}_{tb_n}^*$, $\overline{P}_{fb} = \overline{P}_b^* = \overline{P}_{b_n}$, $n=1,2$.

1) First, we prove $P_{fb} = P_b^*$.

Applying the block matrix inversion formula to (46), the covariance submatrix of the bias is equal to $P_b^*$. That is

$$P_b^{*-1} = [\overline{P}^{*-1}]_b + H_{b_1}^T R_1^{-1} H_{b_1} + H_{b_2}^T R_2^{-1} H_{b_2} - \begin{bmatrix} [\overline{P}^{*-1}]_{tb_1}^T + H_{b_1}^T R_1^{-1} H_{t_1} & [\overline{P}^{*-1}]_{tb_2}^T + H_{b_2}^T R_2^{-1} H_{t_2} \end{bmatrix}$$
$$\times \begin{bmatrix} [\overline{P}^{*-1}]_{t_1} + H_{t_1}^T R_1^{-1} H_{t_1} & O \\ O & [\overline{P}^{*-1}]_{t_2} + H_{t_2}^T R_2^{-1} H_{t_2} \end{bmatrix}^{-1} \begin{bmatrix} [\overline{P}^{*-1}]_{tb_1} + H_{t_1}^T R_1^{-1} H_{b_1} \\ [\overline{P}^{*-1}]_{tb_2} + H_{t_2}^T R_2^{-1} H_{b_2} \end{bmatrix}$$
$$= \overline{P}_b^{*-1} + \overline{P}_b^{*-1}\overline{P}_{tb_1}^{*T}\left(\overline{P}_{t_1}^* - \overline{P}_{tb_1}^*\overline{P}_b^{*-1}\overline{P}_{tb_1}^{*T}\right)^{-1}\overline{P}_{tb_1}^*\overline{P}_b^{*-1} + H_{b_1}^T R_1^{-1} H_{b_1} \quad (62)$$
$$- \left([\overline{P}^{*-1}]_{tb_1}^T + H_{b_1}^T R_1^{-1} H_{t_1}\right)\left([\overline{P}^{*-1}]_{t_1} + H_{t_1}^T R_1^{-1} H_{t_1}\right)^{-1}\left([\overline{P}^{*-1}]_{tb_1} + H_{t_1}^T R_1^{-1} H_{b_1}\right)$$
$$+ \overline{P}_b^{*-1}\overline{P}_{tb_2}^{*T}\left(\overline{P}_{t_2}^* - \overline{P}_{tb_2}^*\overline{P}_b^{*-1}\overline{P}_{tb_2}^{*T}\right)^{-1}\overline{P}_{tb_2}^*\overline{P}_b^{*-1} + H_{b_2}^T R_2^{-1} H_{b_2}$$
$$- \left([\overline{P}^{*-1}]_{tb_2}^T + H_{b_2}^T R_2^{-1} H_{t_2}\right)\left([\overline{P}^{*-1}]_{t_2} + H_{t_2}^T R_2^{-1} H_{t_2}\right)^{-1}\left([\overline{P}^{*-1}]_{tb_2} + H_{t_2}^T R_2^{-1} H_{b_2}\right),$$

where (44) has been applied. Note that the target indices are decoupled in the terms in (62).

In addition, in the single-target KF branch for the target 1 and 2, partitioning $\bar{\boldsymbol{P}}_1^{-1}$ and $\bar{\boldsymbol{P}}_2^{-1}$ according to the target states and sensors biases, we have the following block matrices

$$\bar{\boldsymbol{P}}_1^{-1} = \begin{bmatrix} [\bar{\boldsymbol{P}}_1^{-1}]_{t} & [\bar{\boldsymbol{P}}_1^{-1}]_{tb} \\ [\bar{\boldsymbol{P}}_1^{-1}]_{tb}^{T} & [\bar{\boldsymbol{P}}_1^{-1}]_{b} \end{bmatrix}, \quad (63)$$

$$\bar{\boldsymbol{P}}_2^{-1} = \begin{bmatrix} [\bar{\boldsymbol{P}}_2^{-1}]_{t} & [\bar{\boldsymbol{P}}_2^{-1}]_{tb} \\ [\bar{\boldsymbol{P}}_2^{-1}]_{tb}^{T} & [\bar{\boldsymbol{P}}_2^{-1}]_{b} \end{bmatrix}, \quad (64)$$

For $\bar{\boldsymbol{P}}_1^{-1}$ and $\bar{\boldsymbol{P}}_2^{-1}$, following the similar steps from (42) to (45), we can obtain

$$[\bar{\boldsymbol{P}}_1^{-1}]_{t} = [\bar{\boldsymbol{P}}^{*-1}]_{t_1}, \quad [\bar{\boldsymbol{P}}_1^{-1}]_{tb} = [\bar{\boldsymbol{P}}^{*-1}]_{tb_1}, \quad (65)$$

$$[\bar{\boldsymbol{P}}_2^{-1}]_{t} = [\bar{\boldsymbol{P}}^{*-1}]_{t_2}, \quad [\bar{\boldsymbol{P}}_2^{-1}]_{tb} = [\bar{\boldsymbol{P}}^{*-1}]_{tb_2}, \quad (66)$$

$$[\bar{\boldsymbol{P}}_1^{-1}]_{b} = \bar{\boldsymbol{P}}_b^{*-1} + \bar{\boldsymbol{P}}_b^{*-1} \bar{\boldsymbol{P}}_{tb_1}^{*T} \left( \bar{\boldsymbol{P}}_{t_1}^{*} - \bar{\boldsymbol{P}}_{tb_1}^{*} \bar{\boldsymbol{P}}_b^{*-1} \bar{\boldsymbol{P}}_{tb_1}^{*T} \right)^{-1} \bar{\boldsymbol{P}}_{tb_1}^{*} \bar{\boldsymbol{P}}_b^{*-1}, \quad (67)$$

$$[\bar{\boldsymbol{P}}_2^{-1}]_{b} = \bar{\boldsymbol{P}}_b^{*-1} + \bar{\boldsymbol{P}}_b^{*-1} \bar{\boldsymbol{P}}_{tb_2}^{*T} \left( \bar{\boldsymbol{P}}_{t_2}^{*} - \bar{\boldsymbol{P}}_{tb_2}^{*} \bar{\boldsymbol{P}}_b^{*-1} \bar{\boldsymbol{P}}_{tb_2}^{*T} \right)^{-1} \bar{\boldsymbol{P}}_{tb_2}^{*} \bar{\boldsymbol{P}}_b^{*-1}. \quad (68)$$

Note that only $[\bar{\boldsymbol{P}}_1^{-1}]_b$ and $[\bar{\boldsymbol{P}}_2^{-1}]_b$ are different from the $[\bar{\boldsymbol{P}}^{*-1}]_b$ in $\bar{\boldsymbol{P}}^{*-1}$. $\bar{\boldsymbol{P}}_1^{-1}$ and $\bar{\boldsymbol{P}}_2^{-1}$ can then be rewritten as

$$\bar{\boldsymbol{P}}_1^{-1} = \begin{bmatrix} [\bar{\boldsymbol{P}}^{*-1}]_{t_1} & [\bar{\boldsymbol{P}}^{*-1}]_{tb_1} \\ [\bar{\boldsymbol{P}}^{*-1}]_{tb_1}^{T} & [\bar{\boldsymbol{P}}_1^{-1}]_{b} \end{bmatrix}, \quad (69)$$

$$\bar{\boldsymbol{P}}_2^{-1} = \begin{bmatrix} [\bar{\boldsymbol{P}}^{*-1}]_{t_2} & [\bar{\boldsymbol{P}}^{*-1}]_{tb_2} \\ [\bar{\boldsymbol{P}}^{*-1}]_{tb_2}^{T} & [\bar{\boldsymbol{P}}_2^{-1}]_{b} \end{bmatrix}, \quad (70)$$

According to (22), $\boldsymbol{P}_1^{-1}$ and $\boldsymbol{P}_2^{-1}$ conform to

$$\boldsymbol{P}_1^{-1} = \begin{bmatrix} [\bar{\boldsymbol{P}}^{*-1}]_{t_1} + \boldsymbol{H}_{t_1}^{T} \boldsymbol{R}_1^{-1} \boldsymbol{H}_{t_1} & [\bar{\boldsymbol{P}}^{*-1}]_{tb_1} + \boldsymbol{H}_{t_1}^{T} \boldsymbol{R}_1^{-1} \boldsymbol{H}_{b_1} \\ [\bar{\boldsymbol{P}}^{*-1}]_{tb_1}^{T} + \boldsymbol{H}_{b_1}^{T} \boldsymbol{R}_1^{-1} \boldsymbol{H}_{t_1} & [\bar{\boldsymbol{P}}_1^{-1}]_{b} + \boldsymbol{H}_{b_1}^{T} \boldsymbol{R}_1^{-1} \boldsymbol{H}_{b_1} \end{bmatrix}, \quad (71)$$

$$\boldsymbol{P}_2^{-1} = \begin{bmatrix} [\bar{\boldsymbol{P}}^{*-1}]_{t_2} + \boldsymbol{H}_{t_2}^{T} \boldsymbol{R}_2^{-1} \boldsymbol{H}_{t_2} & [\bar{\boldsymbol{P}}^{*-1}]_{tb_2} + \boldsymbol{H}_{t_2}^{T} \boldsymbol{R}_2^{-1} \boldsymbol{H}_{b_2} \\ [\bar{\boldsymbol{P}}^{*-1}]_{tb_2}^{T} + \boldsymbol{H}_{b_2}^{T} \boldsymbol{R}_2^{-1} \boldsymbol{H}_{t_2} & [\bar{\boldsymbol{P}}_2^{-1}]_{b} + \boldsymbol{H}_{b_2}^{T} \boldsymbol{R}_2^{-1} \boldsymbol{H}_{b_2} \end{bmatrix}. \quad (72)$$

Applying the block matrix inversion formula to (69), (70), (71), and (72), we obtain

$$\boldsymbol{P}_{b_1}^{-1} - \bar{\boldsymbol{P}}_{b_1}^{-1} = \bar{\boldsymbol{P}}_b^{*-1} \bar{\boldsymbol{P}}_{tb_1}^{*T} \left( \bar{\boldsymbol{P}}_{t_1}^{*} - \bar{\boldsymbol{P}}_{tb_1}^{*} \bar{\boldsymbol{P}}_b^{*-1} \bar{\boldsymbol{P}}_{tb_1}^{*T} \right)^{-1} \bar{\boldsymbol{P}}_{tb_1}^{*} \bar{\boldsymbol{P}}_b^{*-1} + \boldsymbol{H}_{b_1}^{T} \boldsymbol{R}_1^{-1} \boldsymbol{H}_{b_1} \\ - \left( [\bar{\boldsymbol{P}}^{*-1}]_{tb_1}^{T} + \boldsymbol{H}_{b_1}^{T} \boldsymbol{R}_1^{-1} \boldsymbol{H}_{t_1} \right) \left( [\bar{\boldsymbol{P}}^{*-1}]_{t_1} + \boldsymbol{H}_{t_1}^{T} \boldsymbol{R}_1^{-1} \boldsymbol{H}_{t_1} \right)^{-1} \left( [\bar{\boldsymbol{P}}^{*-1}]_{tb_1} + \boldsymbol{H}_{t_1}^{T} \boldsymbol{R}_1^{-1} \boldsymbol{H}_{b_1} \right), \quad (73)$$

$$\boldsymbol{P}_{b_2}^{-1} - \bar{\boldsymbol{P}}_{b_2}^{-1} = \bar{\boldsymbol{P}}_b^{*-1} \bar{\boldsymbol{P}}_{tb_2}^{*T} \left( \bar{\boldsymbol{P}}_{t_2}^{*} - \bar{\boldsymbol{P}}_{tb_2}^{*} \bar{\boldsymbol{P}}_b^{*-1} \bar{\boldsymbol{P}}_{tb_2}^{*T} \right)^{-1} \bar{\boldsymbol{P}}_{tb_2}^{*} \bar{\boldsymbol{P}}_b^{*-1} + \boldsymbol{H}_{b_2}^{T} \boldsymbol{R}_2^{-1} \boldsymbol{H}_{b_2} \\ - \left( [\bar{\boldsymbol{P}}^{*-1}]_{tb_2}^{T} + \boldsymbol{H}_{b_2}^{T} \boldsymbol{R}_2^{-1} \boldsymbol{H}_{t_2} \right) \left( [\bar{\boldsymbol{P}}^{*-1}]_{t_2} + \boldsymbol{H}_{t_2}^{T} \boldsymbol{R}_2^{-1} \boldsymbol{H}_{t_2} \right)^{-1} \left( [\bar{\boldsymbol{P}}^{*-1}]_{tb_2} + \boldsymbol{H}_{t_2}^{T} \boldsymbol{R}_2^{-1} \boldsymbol{H}_{b_2} \right), \quad (74)$$

where (67) and (68) have been applied.

Substituting (73) and (74) into (62), we obtain

$$\boldsymbol{P}_b^{*-1} - \bar{\boldsymbol{P}}_b^{*-1} = \boldsymbol{P}_{b_1}^{-1} - \bar{\boldsymbol{P}}_{b_1}^{-1} + \boldsymbol{P}_{b_2}^{-1} - \bar{\boldsymbol{P}}_{b_2}^{-1}. \quad (75)$$

Given $\bar{\boldsymbol{P}}_b^{*-1} = \bar{\boldsymbol{P}}_{fb}^{-1}$, comparing (75) with (26), we draw the conclusion

$$\boldsymbol{P}_b^{*-1} = \bar{\boldsymbol{P}}_{fb}^{-1} + \boldsymbol{P}_{b_1}^{-1} - \bar{\boldsymbol{P}}_{b_1}^{-1} + \boldsymbol{P}_{b_2}^{-1} - \bar{\boldsymbol{P}}_{b_2}^{-1} = \boldsymbol{P}_{fb}^{-1}. \quad (76)$$

2) Second, we prove $\hat{\boldsymbol{b}}_f = \hat{\boldsymbol{b}}^*$ and $\hat{\boldsymbol{x}}_{ft_n} = \hat{\boldsymbol{x}}_{t_n}^*$.

Applying the block matrix inversion formula to (71) and (72), we obtain

$$\boldsymbol{P}_{tb_1} \boldsymbol{P}_{b_1}^{-1} = -\left( [\bar{\boldsymbol{P}}^{*-1}]_{t_1} + \boldsymbol{H}_{t_1}^{T} \boldsymbol{R}_1^{-1} \boldsymbol{H}_{t_1} \right)^{-1} \left( [\bar{\boldsymbol{P}}^{*-1}]_{tb_1} + \boldsymbol{H}_{t_1}^{T} \boldsymbol{R}_1^{-1} \boldsymbol{H}_{b_1} \right), \quad (77)$$

$$\boldsymbol{P}_{tb_2} \boldsymbol{P}_{b_2}^{-1} = -\left( [\bar{\boldsymbol{P}}^{*-1}]_{t_2} + \boldsymbol{H}_{t_2}^{T} \boldsymbol{R}_2^{-1} \boldsymbol{H}_{t_2} \right)^{-1} \left( [\bar{\boldsymbol{P}}^{*-1}]_{tb_2} + \boldsymbol{H}_{t_2}^{T} \boldsymbol{R}_2^{-1} \boldsymbol{H}_{b_2} \right), \quad (78)$$

$$\boldsymbol{P}_{b_1}^{-1} - \left( [\bar{\boldsymbol{P}}_1^{-1}]_b + \boldsymbol{H}_{b_1}^{T} \boldsymbol{R}_1^{-1} \boldsymbol{H}_{b_1} \right) = -\left( [\bar{\boldsymbol{P}}^{*-1}]_{tb_1}^{T} + \boldsymbol{H}_{b_1}^{T} \boldsymbol{R}_1^{-1} \boldsymbol{H}_{t_1} \right)\left( [\bar{\boldsymbol{P}}^{*-1}]_{t_1} + \boldsymbol{H}_{t_1}^{T} \boldsymbol{R}_1^{-1} \boldsymbol{H}_{t_1} \right)^{-1} \left( [\bar{\boldsymbol{P}}^{*-1}]_{tb_1} + \boldsymbol{H}_{t_1}^{T} \boldsymbol{R}_1^{-1} \boldsymbol{H}_{b_1} \right) \\ = \left( [\bar{\boldsymbol{P}}^{*-1}]_{tb_1}^{T} + \boldsymbol{H}_{b_1}^{T} \boldsymbol{R}_1^{-1} \boldsymbol{H}_{t_1} \right) \boldsymbol{P}_{tb_1} \boldsymbol{P}_{b_1}^{-1}, \quad (79)$$

$$\boldsymbol{P}_{b_2}^{-1} - \left( [\bar{\boldsymbol{P}}_2^{-1}]_b + \boldsymbol{H}_{b_2}^{T} \boldsymbol{R}_2^{-1} \boldsymbol{H}_{b_2} \right) = -\left( [\bar{\boldsymbol{P}}^{*-1}]_{tb_2}^{T} + \boldsymbol{H}_{b_2}^{T} \boldsymbol{R}_2^{-1} \boldsymbol{H}_{t_2} \right)\left( [\bar{\boldsymbol{P}}^{*-1}]_{t_2} + \boldsymbol{H}_{t_2}^{T} \boldsymbol{R}_2^{-1} \boldsymbol{H}_{t_2} \right)^{-1} \left( [\bar{\boldsymbol{P}}^{*-1}]_{tb_2} + \boldsymbol{H}_{t_2}^{T} \boldsymbol{R}_2^{-1} \boldsymbol{H}_{b_2} \right) \\ = \left( [\bar{\boldsymbol{P}}^{*-1}]_{tb_2}^{T} + \boldsymbol{H}_{b_2}^{T} \boldsymbol{R}_2^{-1} \boldsymbol{H}_{t_2} \right) \boldsymbol{P}_{tb_2} \boldsymbol{P}_{b_2}^{-1}. \quad (80)$$

Substituting (45) and (46) into (12) and expanding it, we obtain

$$\left( [\bar{\boldsymbol{P}}^{*-1}]_{t_1} + \boldsymbol{H}_{t_1}^{T} \boldsymbol{R}_1^{-1} \boldsymbol{H}_{t_1} \right) \hat{\boldsymbol{x}}_{t_1}^{*} + \left( [\bar{\boldsymbol{P}}^{*-1}]_{tb_1} + \boldsymbol{H}_{t_1}^{T} \boldsymbol{R}_1^{-1} \boldsymbol{H}_{b_1} \right) \hat{\boldsymbol{b}}^{*} = [\bar{\boldsymbol{P}}^{*-1}]_{t_1} \hat{\boldsymbol{x}}_{t_1}^{*}(k+1|k) + [\bar{\boldsymbol{P}}^{*-1}]_{tb_1} \hat{\boldsymbol{b}}^{*}(k+1|k) + \boldsymbol{H}_{t_1}^{T} \boldsymbol{R}_1^{-1} \boldsymbol{z}_1, \quad (81)$$

$$\left( [\bar{\boldsymbol{P}}^{*-1}]_{t_2} + \boldsymbol{H}_{t_2}^{T} \boldsymbol{R}_2^{-1} \boldsymbol{H}_{t_2} \right) \hat{\boldsymbol{x}}_{t_2}^{*} + \left( [\bar{\boldsymbol{P}}^{*-1}]_{tb_2} + \boldsymbol{H}_{t_2}^{T} \boldsymbol{R}_2^{-1} \boldsymbol{H}_{b_2} \right) \hat{\boldsymbol{b}}^{*} = [\bar{\boldsymbol{P}}^{*-1}]_{t_2} \hat{\boldsymbol{x}}_{t_2}^{*}(k+1|k) + [\bar{\boldsymbol{P}}^{*-1}]_{tb_2} \hat{\boldsymbol{b}}^{*}(k+1|k) + \boldsymbol{H}_{t_2}^{T} \boldsymbol{R}_2^{-1} \boldsymbol{z}_2, \quad (82)$$

$$\left( [\bar{\boldsymbol{P}}^{*-1}]_{tb_1}^{T} + \boldsymbol{H}_{b_1}^{T} \boldsymbol{R}_1^{-1} \boldsymbol{H}_{t_1} \right) \hat{\boldsymbol{x}}_{t_1}^{*} + \left( [\bar{\boldsymbol{P}}^{*-1}]_{tb_2}^{T} + \boldsymbol{H}_{b_2}^{T} \boldsymbol{R}_2^{-1} \boldsymbol{H}_{t_2} \right) \hat{\boldsymbol{x}}_{t_2}^{*} + \left( [\bar{\boldsymbol{P}}^{*-1}]_b + \boldsymbol{H}_{b_1}^{T} \boldsymbol{R}_1^{-1} \boldsymbol{H}_{b_1} + \boldsymbol{H}_{b_2}^{T} \boldsymbol{R}_2^{-1} \boldsymbol{H}_{b_2} \right) \hat{\boldsymbol{b}}^{*} \\ = [\bar{\boldsymbol{P}}^{*-1}]_{tb_1}^{T} \hat{\boldsymbol{x}}_{t_1}^{*}(k+1|k) + [\bar{\boldsymbol{P}}^{*-1}]_{tb_2}^{T} \hat{\boldsymbol{x}}_{t_2}^{*}(k+1|k) + [\bar{\boldsymbol{P}}^{*-1}]_b \hat{\boldsymbol{b}}^{*}(k+1|k) + \boldsymbol{H}_{b_1}^{T} \boldsymbol{R}_1^{-1} \boldsymbol{z}_1 + \boldsymbol{H}_{b_2}^{T} \boldsymbol{R}_2^{-1} \boldsymbol{z}_2. \quad (83)$$

Likewise, substituting (69), (70), (71), and (72) into (21) and expanding them, we obtain

$$\left([\overline{P}^{*-1}]_{t_1} + H_{t_1}^T R_1^{-1} H_{t_1}\right)\hat{x}_{t_1} + \left([\overline{P}^{*-1}]_{tb_1} + H_{t_1}^T R_1^{-1} H_{b_1}\right)\hat{b}_1 = [\overline{P}^{*-1}]_{t_1}\hat{x}_{t_1}(k+1|k) + [\overline{P}^{*-1}]_{tb_1}\hat{b}_1(k+1|k) + H_{t_1}^T R_1^{-1} z_1, \quad (84)$$

$$\left([\overline{P}^{*-1}]_{tb_1}^T + H_{b_1}^T R_1^{-1} H_{t_1}\right)\hat{x}_{t_1} + \left([\overline{P}_1^{-1}]_b + H_{b_1}^T R_1^{-1} H_{b_1}\right)\hat{b}_1 = [\overline{P}^{*-1}]_{tb_1}^T \hat{x}_{t_1}(k+1|k) + [\overline{P}_1^{-1}]_b \hat{b}_1(k+1|k) + H_{b_1}^T R_1^{-1} z_1, \quad (85)$$

$$\left([\overline{P}^{*-1}]_{t_2} + H_{t_2}^T R_2^{-1} H_{t_2}\right)\hat{x}_{t_2} + \left([\overline{P}^{*-1}]_{tb_2} + H_{t_2}^T R_2^{-1} H_{b_2}\right)\hat{b}_2 = [\overline{P}^{*-1}]_{t_2}\hat{x}_{t_2}(k+1|k) + [\overline{P}^{*-1}]_{tb_2}\hat{b}_2(k+1|k) + H_{t_2}^T R_2^{-1} z_2, \quad (86)$$

$$\left([\overline{P}^{*-1}]_{tb_2}^T + H_{b_2}^T R_2^{-1} H_{t_2}\right)\hat{x}_{t_2} + \left([\overline{P}_2^{-1}]_b + H_{b_2}^T R_2^{-1} H_{b_2}\right)\hat{b}_2 = [\overline{P}^{*-1}]_{tb_2}^T \hat{x}_{t_2}(k+1|k) + [\overline{P}_2^{-1}]_b \hat{b}_2(k+1|k) + H_{b_2}^T R_2^{-1} z_2. \quad (87)$$

Subtracting (84) from (81), we have

$$\left([\overline{P}^{*-1}]_{t_1} + H_{t_1}^T R_1^{-1} H_{t_1}\right)\hat{x}_{t_1}^* + \left([\overline{P}^{*-1}]_{tb_1} + H_{t_1}^T R_1^{-1} H_{b_1}\right)\hat{b}^* = \left([\overline{P}^{*-1}]_{t_1} + H_{t_1}^T R_1^{-1} H_{t_1}\right)\hat{x}_{t_1} + \left([\overline{P}^{*-1}]_{tb_1} + H_{t_1}^T R_1^{-1} H_{b_1}\right)\hat{b}_1. \quad (88)$$

Then we can obtain

$$\hat{x}_{t_1}^* = \hat{x}_{t_1} - \left([\overline{P}^{*-1}]_{t_1} + H_{t_1}^T R_1^{-1} H_{t_1}\right)^{-1}\left([\overline{P}^{*-1}]_{tb_1} + H_{t_1}^T R_1^{-1} H_{b_1}\right)(\hat{b}^* - \hat{b}_1) = \hat{x}_{t_1} + P_{tb_1} P_{b_1}^{-1}(\hat{b}^* - \hat{b}_1), \quad (89)$$

where (77) has been applied.

Likewise, subtracting (86) from (82), we can obtain

$$\hat{x}_{t_2}^* = \hat{x}_{t_2} + P_{tb_2} P_{b_2}^{-1}(\hat{b}^* - \hat{b}_2). \quad (90)$$

Subtracting (85) and (87) from (83), we obtain

$$\left([\overline{P}^{*-1}]_{tb_1}^T + H_{b_1}^T R_1^{-1} H_{t_1}\right)(\hat{x}_{t_1}^* - \hat{x}_{t_1}) + \left([\overline{P}^{*-1}]_{tb_2}^T + H_{b_2}^T R_2^{-1} H_{t_2}\right)(\hat{x}_{t_2}^* - \hat{x}_{t_2}) + \left([\overline{P}^{*-1}]_b + H_{b_1}^T R_1^{-1} H_{b_1} + H_{b_2}^T R_2^{-1} H_{b_2}\right)\hat{b}^* - [\overline{P}^{*-1}]_b \hat{b}^*(k+1|k)$$
$$= \left([\overline{P}_1^{-1}]_b + H_{b_1}^T R_1^{-1} H_{b_1}\right)\hat{b}_1 - [\overline{P}_1^{-1}]_b \hat{b}_1(k+1|k) + \left([\overline{P}_2^{-1}]_b + H_{b_2}^T R_2^{-1} H_{b_2}\right)\hat{b}_2 - [\overline{P}_2^{-1}]_b \hat{b}_2(k+1|k). \quad (91)$$

Substituting (89) and (90) into (91) and applying (79) and (80), we can obtain

$$\left[P_{b_1}^{-1} - \left([\overline{P}_1^{-1}]_b + H_{b_1}^T R_1^{-1} H_{b_1}\right)\right](\hat{b}^* - \hat{b}_1) + \left[P_{b_2}^{-1} - \left([\overline{P}_2^{-1}]_b + H_{b_2}^T R_2^{-1} H_{b_2}\right)\right](\hat{b}^* - \hat{b}_2)$$
$$+ \left([\overline{P}^{*-1}]_b + H_{b_1}^T R_1^{-1} H_{b_1} + H_{b_2}^T R_2^{-1} H_{b_2}\right)\hat{b}^* - [\overline{P}^{*-1}]_b \hat{b}^*(k+1|k) \quad (92)$$
$$= \left([\overline{P}_1^{-1}]_b + H_{b_1}^T R_1^{-1} H_{b_1}\right)\hat{b}_1 - [\overline{P}_1^{-1}]_b \hat{b}_1(k+1|k) + \left([\overline{P}_2^{-1}]_b + H_{b_2}^T R_2^{-1} H_{b_2}\right)\hat{b}_2 - [\overline{P}_2^{-1}]_b \hat{b}_2(k+1|k).$$

Merging the same terms in (92), we obtain

$$\left(P_{b_1}^{-1} + P_{b_2}^{-1} + [\overline{P}^{*-1}]_b - [\overline{P}_1^{-1}]_b - [\overline{P}_2^{-1}]_b\right)\hat{b}^* - [\overline{P}^{*-1}]_b \hat{b}^*(k+1|k) = P_{b_1}^{-1}\hat{b}_1 - [\overline{P}_1^{-1}]_b \hat{b}_1(k+1|k) + P_{b_2}^{-1}\hat{b}_2 - [\overline{P}_2^{-1}]_b \hat{b}_2(k+1|k). \quad (93)$$

Applying $\hat{b}^*(k+1|k) = \hat{b}_1(k+1|k) = \hat{b}_2(k+1|k)$, $\overline{P}_b^{*-1} = \overline{P}_{b_1}^{-1} = \overline{P}_{b_2}^{-1}$, and substituting (44), (67) and (68) into (93), we obtain

$$\left(P_{b_1}^{-1} + P_{b_2}^{-1} - \overline{P}_b^{*-1}\right)\hat{b}^* - \overline{P}_b^{*-1}\hat{b}^*(k+1|k)$$
$$= P_{b_1}^{-1}\hat{b}_1 - \overline{P}_{b_1}^{-1}\hat{b}_1(k+1|k) + P_{b_2}^{-1}\hat{b}_2 - \overline{P}_{b_2}^{-1}\hat{b}_2(k+1|k). \quad (94)$$

According to (76), it can be readily obtained that $P_{b_1}^{-1} + P_{b_2}^{-1} - \overline{P}_b^{*-1} = P_{fb}^{-1}$. Comparing (94) with (26), we draw the conclusion

$$\hat{b}^* = \hat{b}_f. \quad (95)$$

Substituting (95) into (89) and (90) and comparing them with (27), we draw the conclusion

$$\hat{x}_{t_n}^* = \hat{x}_{ft_n}, n = 1, 2. \quad (96)$$

3) Finally, we prove $P_{ft_n} = P_{t_n}^*$ and $P_{ftb_n} = P_{tb_n}^*$.

Note that (48) can also be expressed as

$$\begin{bmatrix} P_{tb_1}^* \\ P_{tb_2}^* \end{bmatrix} = \begin{bmatrix} -\left([\overline{P}^{*-1}]_{t_1} + H_{t_1}^T R_1^{-1} H_{t_1}\right)^{-1}\left([\overline{P}^{*-1}]_{tb_1} + H_{t_1}^T R_1^{-1} H_{b_1}\right) \\ -\left([\overline{P}^{*-1}]_{t_2} + H_{t_2}^T R_2^{-1} H_{t_2}\right)^{-1}\left([\overline{P}^{*-1}]_{tb_2} + H_{t_2}^T R_2^{-1} H_{b_2}\right) \end{bmatrix} P_b^* = \begin{bmatrix} P_{tb_1} P_{b_1}^{-1} P_b^* \\ P_{tb_2} P_{b_2}^{-1} P_b^* \end{bmatrix}, \quad (97)$$

where (77) and (78) has been applied.

Comparing (97) with (28) and applying (76) (i.e. $P_b^* = P_{fb}$), we draw the conclusion

$$P_{tb_n}^* = P_{ftb_n}, n = 1, 2. \quad (98)$$

Besides, according to (28), (29), (71) and (72), we obtain

$$P_{fn}^{-1} = \begin{bmatrix} P_{ft_n} & P_{ftb_n} \\ P_{ftb_n}^T & P_{fb} \end{bmatrix}^{-1} = \begin{bmatrix} [\overline{P}^{*-1}]_{t_n} + H_{t_n}^T R_n^{-1} H_{t_n} & [\overline{P}^{*-1}]_{tb_n} + H_{t_n}^T R_n^{-1} H_{b_n} \\ [\overline{P}^{*-1}]_{tb_n}^T + H_{b_n}^T R_n^{-1} H_{t_n} & [\overline{P}_n^{-1}]_b + H_{b_n}^T R_n^{-1} H_{b_n} + P_{fb}^{-1} - P_{b_n}^{-1} \end{bmatrix}, n = 1, 2. \quad (99)$$

Applying the block matrix inversion formula to (99), we obtain

$$\boldsymbol{P}_{\text{ft}_n} - \boldsymbol{P}_{\text{ftb}_n}\boldsymbol{P}_{\text{fb}}^{-1}\boldsymbol{P}_{\text{ftb}_n}^{\text{T}} = \left(\left[\bar{\boldsymbol{P}}^{*-1}\right]_{\text{t}_n} + \boldsymbol{H}_{\text{t}_n}^{\text{T}}\boldsymbol{R}_n^{-1}\boldsymbol{H}_{\text{t}_n}\right)^{-1}, n=1,2. \quad (100)$$

Comparing (100) with (49) and applying (76) (i.e. $\boldsymbol{P}_{\text{b}}^* = \boldsymbol{P}_{\text{fb}}$) and (98) (i.e. $\boldsymbol{P}_{\text{tb}_n}^* = \boldsymbol{P}_{\text{ftb}_n}$), we draw the conclusion

$$\boldsymbol{P}_{\text{t}_n}^* = \boldsymbol{P}_{\text{ft}_n}, n=1,2. \quad (101)$$

In summary, eq. (76), (95), (96), (98), and (101) give the desired conclusions at $k+1$. The theorem is proved when $N=2$.

ACKNOWLEDGMENT

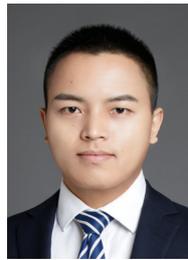

**Jianxin Yi** (M'18) received the B.E. degree in electrical and electronic engineering, in 2011, and the Ph.D. degree in radio physics, in 2016, both from Wuhan University, China. From Aug. 2014 to Aug. 2015, he was also a visiting Ph.D. student at University of Calgary, Canada.

He is now a research associate professor at the School of Electronic Information, Wuhan University. He was the recipient of the 2017 Excellent Doctoral Dissertation Award of the Chinese Institute of Electronics. He has been supported by the Postdoctoral Innovation Talent Support Program of China. His main research interests include radar signal processing, target tracking, and information fusion.

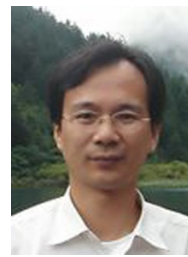

**Xianrong Wan** received the B.E. degree from the former Wuhan Technical University of Surveying and Mapping, China, in 1997, and the Ph.D. degree from Wuhan University, China, in 2005.

He is now a professor and Ph.D. candidate supervisor of the School of Electronic Information, Wuhan University. Recent

years he has hosted and participated in more than ten national research projects, and published more than 80 academic papers. His main research interests include design of new radar system such as passive radar, over-the-horizon radar, and array signal processing.

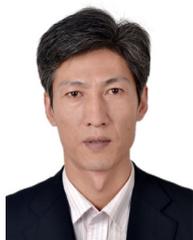

**Deshi Li** received the Ph.D. degree in Computer Application Technology from Wuhan University. From Oct. 2006 to Oct. 2007, he was a visiting scholar of the Network Lab of the University of California at Davis.

He is now a professor and the dean of the School of Electronic Information, Wuhan University. He currently serves as a member of the Internet of Things Expert Committee and a member of the Education Committee of Chinese Institute of Electronics. He is the Associate Chief scientist in Space Communication area of Collaborative Innovation Center of Geospatial Technology, and is also an Executive Trustee member of China Cloud System Pioneer Strategic Alliance. His main research interests include sensor networks, intelligence system, SOC design and verification methodology.